\documentclass[epj]{svjour}

\usepackage[latin1]{inputenc}

\usepackage{graphicx}
\usepackage{color}

\usepackage{amssymb}
\usepackage{color}%
\usepackage{booktabs}

\begin{document}

\title{Second virial coefficients of light  nuclear  clusters 
and their chemical freeze-out in nuclear collisions}
\titlerunning{Second virial coefficients of light  nuclear  clusters ...}



\author{K. A. Bugaev\inst{1, 2}
\thanks{\email{bugaev@th.physik.uni-frankfurt.de}},
O. V.  Vitiuk\inst{2, 3},
B. E. Grinyuk\inst{1},
V. V. Sagun\inst{1, 4}, 
N. S. Yakovenko\inst{2},
O. I. Ivanytskyi\inst{1, 4},  
G. M. Zinovjev\inst{1}, 
D. B. Blaschke\inst{5, 6, 7}
E. G. Nikonov\inst{8},
L. V.  Bravina\inst{3}, 
E.  E.  Zabrodin\inst{3, 9}, 
S. Kabana\inst{10}, 
S. V. Kuleshov\inst{11},
G. R.  Farrar\inst{12},
E. S. Zherebtsova\inst{7, 13}
 and A. V.  Taranenko\inst{7}\\
}
\authorrunning{K. A. Bugaev et al.}
\institute{                    
  \inst{1} Bogolyubov Institute for Theoretical Physics,
Metrologichna str. 14$^B$, Kyiv 03680, Ukraine\\
  \inst{2} Department of Physics, Taras Shevchenko National University of Kyiv, 03022 Kyiv, Ukraine\\
   \inst{3} University of Oslo, POB 1048 Blindern, N-0316 Oslo, Norway\\  
  \inst{4} CFisUC, Department of Physics, University of Coimbra, 3004-516 Coimbra, Portugal\\
    \inst{5} Institute of Theoretical Physics, University of Wroclaw, Max Born Pl. 9, 50-204 Wroclaw, Poland\\
   \inst{6} Bogoliubov Laboratory of Theoretical Physics, JINR Dubna, Joliot-Curie Str. 6, 141980 Dubna, Russia\\
   \inst{7} National Research Nuclear University (MEPhI), Kashirskoe Shosse 31, 115409 Moscow, Russia\\
  \inst{8} Laboratory for Information Technologies, Joint Institute for Nuclear Research,   Dubna 141980, Russia\\
   \inst{9} Skobeltsyn Institute of Nuclear Physics, Moscow State University, 119899 Moscow, Russia\\
      \inst{10} Instituto de Alta Investigaci\'on, Universidad de Tarapac\'a, Casilla 7D, Arica, Chile\\
      \inst{11} Departamento de Ciencias  F\'{\i}sicas, Universidad Andres Bello, Sazi\'e 2212, Piso 7, Santiago, Chile\\
    \inst{12}  Department of Physics, New York University, 
    New York, NY 10003, USA\\
    \inst{13} {Institute for Nuclear Research, Russian Academy of Science, 108840 Moscow, Russia}
}
\abstract{Here we develop a new strategy to analyze the chemical freeze-out of light (anti)nuclei 
produced in  high energy collisions of heavy atomic nuclei within an advanced version of the hadron resonance gas model. 
It is based on two different, but complementary  approaches to model the hard-core repulsion between the light nuclei and hadrons. The first approach  is based on an approximate treatment of the equivalent hard-core radius of a roomy nuclear cluster and pions, while the second approach is rigorously derived here using  a self-consistent treatment of classical excluded  volumes of light (anti)nuclei  and hadrons.   
By construction, in a hadronic medium dominated by pions,  both approaches should give the same results.
Employing this strategy to the analysis of  hadronic and light (anti)nuclei multiplicities measured by  ALICE   at 
$\sqrt{s_{NN}} =2.76$ TeV and by STAR  at $\sqrt{s_{NN}} =200$ GeV, we  got rid of  the  existing ambiguity in the  description of light (anti)nuclei data and determined the  chemical freeze-out  parameters of  nuclei with high accuracy and confidence. 
At ALICE energy the nuclei are frozen  prior to the hadrons at the temperature $T = 175.1^{+2.3}_{-3.9}$ MeV, 
while at STAR energy there is a single freeze-out of hadrons and nuclei at the temperature $T = 167.2 \pm 3.9$ MeV.
We argue that the found chemical freeze-out volumes of nuclei  can be considered as  the volumes of quark-gluon bags 
that  produce the nuclei at the moment of hadronization. 
}
\maketitle

\section{Introduction} \label{Intro}

The concept of  hard-core repulsion plays an important role in the statistical mechanics of classical systems  since, despite its simplicity,
it allows  one  to correctly reproduce the basic properties of real gases at short distances.
Its importance  in  describing the multiplicities of hadrons produced in the central high energy nuclear (A+A) collisions is beyond any doubts. 
In atomic physics it is clear that the hard core in the intermolecular interaction has its fundamental origin in the Pauli exclusion principle 
and the electron exchange correlations between atoms and molecules (see, e.g., \cite{Ebeling:2008mg} and references therein) which allows {one}
to predict the composition and thermodynamics of inertial fusion plasmas due to the account of the Pauli-blocking effect between atomic clusters
\cite{Ropke:2018ewt}. 
{However,} the application of the fundamental Pauli principle on the quark level to account for a repulsive hard core in the 
interaction among hadrons is still in its infancy \cite{Blaschke:2020qrs}.
The account of the Pauli blocking effect for light clusters in nuclear matter is meanwhile well-elaborated for not too high temperatures
\cite{Typel:2009sy} and for applications to the composition of supernova matter \cite{Hempel:2015yma,Ropke:2020peo}, where usually excluded
volume approaches are applied to account for light cluster abundances \cite{Lattimer:1991nc,Shen:1998gq,Hempel:2011kh}.
Within the quantum statistical approach, the second virial coefficient is addressed via a generalized Beth-Uhlenbeck equation of state which accounts {for}
medium effects on the scattering phase shifts among clusters (cluster virial expansion \cite{Ropke:2012qv}).
The latter not only  describes systematically the in-medium modification of the hard-core interaction, but ultimately leads to the Mott dissociation of the nuclear 
clusters due to Pauli blocking.
The generalization of this successful quantum statistical approach to the higher temperatures by including all species of a hadron resonance gas 
and the treatment of repulsive Pauli-blocking effects on the basis of their fermionic quark substructure is a formidable task that has just been started 
\cite{KAB_Phi-approach3}. 
For the time being, one can already get interesting insights for the discussion of chemical freeze-out (CFO) of light clusters in the QCD phase diagram,
in the context of ongoing discussions of the puzzle why these clusters freeze out in ultrarelativistic heavy-ion collisions at CERN and BNL 
according to predictions of the thermal statistical model at the same high temperature $T_{CFO} \approx 160$ MeV like all the other hadrons while 
their binding energies are at least an order of magnitude smaller. 

{
This puzzle of the light nuclei production at LHC and RHIC has been discussed in many recent papers 
\cite{Mrowczynski:2016xqm,Sun:2017xrx,Sun:2018jhg,Sun:2018mqq,Vovchenko:2018fiy,Bellini:2018epz,Bellini:2020cbj,Cai:2019jtk,Aichelin:2019tnk,Vitiyuk_Ref2,Vitiyuk_Ref3}, from both alternative points of view: the coalescence of nucleons (and hyperons, if applicable) in the final state after thermal freezeout of the hadrons on the one hand, and the CFO 
of the nuclei according to the thermal statistical model together with the other hadronic species directly in the vicinity of the hadronization transition in the QCD phase diagram on the other.
}

When drawing the lines for the Mott dissociation of light clusters as derived from the quantum statistical model into the QCD phase diagram one observes
\cite{Blaschke:2020jmv,Blaschke:2020gqr}
that at the conditions of LHC and STAR experiments the medium modifications for nuclear clusters are not important, so that they can be expected to
follow the ordinary thermal statistical model albeit including a hard core repulsion as in free space.
Therefore, we devote the present work to { extending}  the concept of hard core repulsion for hadronic systems with nuclear clusters in a thermodynamically consistent way and will apply it to the description of hadron and light cluster yields obtained in these experiments.

The real breakthrough in achieving a very high accuracy in {the} description of  hadronic yields 
measured from the low AGS BNL collision energy ($\sqrt{s_{NN}} =2.7$ GeV) to the LHC CERN one  ($\sqrt{s_{NN}} =2.76$ TeV) is related to the hadron resonance gas model (HRGM) with several hard-core radii of hadronic species \cite{MHRGM1,MHRGM2,MHRGM3,MHRGM4}, i.e. with the multicomponent hard-core  repulsion.  
Indeed, using just two extra  parameters,
the hard-core radii of pions $R_\pi$ and kaons $R_K$, in addition  to the  hard-core radii of baryons $R_b$ and the ones of other mesons $R_m$ which are traditionally employed  in the HRGM, it was possible to achieve a very accurate description of all independent  hadron multiplicity ratios measured prior to the LHC era with a $\chi^2/dof$ which is  in the range 
between 1.15 \cite{MHRGM2,MHRGM3,MHRGM4} and 0.96 \cite{Sagun14}.
The high accuracy  achieved by the HRGM with multicomponent hard-core repulsion 
allowed us not only to elucidate the  characteristics  of the CFO of  A+A collisions, but also to 
reveal new irregularities of thermodynamic quantities at the CFO and to formulate new signals of two phase transitions
\cite{GSA15,GSA16,GSA16b,Signals18,Signals19}  which are expected to exist  in strongly interacting matter.  

We have to remind that traditionally the CFO is defined as the moment after which the inelastic reactions stop to exist, while  the evolution of hadronic matter  
is dominated  by  elastic reactions towards thermal freeze-out and decays of resonances \cite{PBM06}. 
 
However, the multicomponent versions of the HRGM based on the Van der Waals (VdW) approximation to the hard-core repulsion, i.e. which employ  the classical second virial coefficients,  are rather complicated and they take a lot of CPU time, since for  $N$ different hard-core radii  for each iteration of the fitting process of  experimental data  
one has to solve the  system of $N+1$ transcendental equations which involve hundreds of double integrals \cite{MHRGM1,MHRGM2,MHRGM3,MHRGM4,Sagun14}. Therefore, the application of the multicomponent HRGM based on VdW approximation to cases of $N \gg 1$ is rather problematic. Fortunately, an entirely new and efficient   approach 
to treat the multicomponent hard-core repulsion for   large values of $N$ was invented in Ref.  \cite{IST1}. 
This  novel approach based on  the induced surface tension  concept  has two important advantages over the other 
multicomponent versions of the HRGM: first, the number of equations to be solved is always two and it does not depend on $N$ and, second, it allows one to go beyond the VdW approximation \cite{IST2,IST3,QSTAT2019,Nazar2019}.
Note that the classical virial coefficients are traditionally denoted as the excluded volumes (per particle).

Despite the great achievements of the HRGM one important problem of the CFO was not resolved until recently. 
It is the CFO  puzzle of   light (anti)nuclei yields  measured   by 
the STAR RHIC collaboration in Au+Au central collisions  \cite{STARA1,STARA2,STARA3} at the center-of-mass collision energy 
$\sqrt{s_{NN}} =200$ GeV and the ones obtained recently   by 
the ALICE CERN  collaboration in Pb+Pb collisions at the center-of-mass collision energy $\sqrt{s_{NN}} =2.76$ TeV \cite{KAB_Ref1a,KAB_Ref1b,KAB_Ref1c}. 

There are many important aspects of the CFO  puzzle of  light (anti)nuclei yields 
\cite{Cai:2019jtk,Vitiyuk_Ref2,Vitiyuk_Ref3,Edward2018,Shuryak:2019ikv,KAB_Jean,KAB_Ref2,PBM18,KAB_Ref3,PBM19,Grinyuk2020} 
measured in A+A collisions,  
{ 
but  in our opinion the central one,
} 
is  the value of their  CFO temperature $T_A$. 
This is so, since without the reliable 
knowledge of their  CFO temperature $T_A$ one cannot formulate a physically adequate model for the production of  deuterons (d),  helium-3 ($^3$He),  helium-4 ($^4$He) and hyper-triton ($^3_\Lambda$H) and their antiparticles 
in A+A collisions and a model of their thermalization as well.  
{
Other approaches in the literature  which describe the production of nuclei in heavy ion collisions obtain estimates for $T_A$ using  extensions of the HRGM that consider the nuclei as point-like particles \cite{KAB_Jean} or assume the hard-core radius of all light (anti)nuclei to be equal to the hard-core radius of baryons  $R_b$ \cite{KAB_Ref2}.  It is of interest to go a step further and aim to describe nuclei in a more realistic way.

In our previous work \cite{KAB_Ref3}  a more elaborate HRGM has been presented that is based on the concept of induced surface tension \cite{IST1,IST2,IST3}. It uses an approximate expression for the hard-core radius of light (anti)nuclei denoted as bag model approximation (see below). This   restriction was overcome recently in  \cite{Grinyuk2020}, where it was 
shown  that the bag model approximation can safely be used for pion-dominated matter. However,  the derivation of the equation of state (EoS) which extends  the induced surface tension  concept  to  the classical second virial coefficients of light (anti)nuclei as suggested in [40] is a heuristic one. 

In the present work we develop a mathematically rigorous treatment of a mixture of hadrons and light (anti)nuclei with hard-core repulsion based on the induced surface tension concept  \cite{IST1,IST2,IST3}.  In addition, with the help of this newly developed HRGM we analyze here not only the ALICE $\sqrt{s_{NN}} =2.76$ TeV data   on light (anti)nuclei   \cite{KAB_Ref1a,KAB_Ref1b,KAB_Ref1c}, but also  the STAR $\sqrt{s_{NN}} =200$ GeV data \cite{STARA1,STARA2,STARA3}.  
\color{blue} Our  experience on achieving  the  accurate  description of the hadronic data  documented  in  Refs. 
\cite{MHRGM1,MHRGM2,MHRGM3,MHRGM4,Sagun14,GSA15,GSA16,Signals18,Signals19,IST2,IST3} gives us a confidence that an essential  improvement of the light nuclei data description  will  help the community to resolve the puzzles of the   CFO of light nuclei. 
}

The work is organized as follows. In Sect.~2 the mathematically rigorous derivation of the induced surface tension EoS 
for the mixture of hadrons and nuclei with classical second virial coefficients is presented. Sect.~3 contains the results on 
two models of the CFO of light (anti)nuclei produced in the central A+A collisions on LHC and RHIC.
Sect.~4 is devoted to the discussion of the obtained results and summarizes our conclusions.

\section{Self-Consistent Treatment of Classical Excluded Volumes} \label{Sect2}

In this section,  we briefly show how to extend the method of self-consistent treatment of classical systems with multicomponent hard-core interaction to the case of  interaction of hadrons and light nuclei.  It was introduced in \cite{QSTAT2019}
and successfully applied in \cite{Nazar2019} to mixtures of classical  hard spheres  and hard discs of different sizes.

{
There are  three major reasons   to consider  the HRGM with multicomponent hard-core 
repulsion  as the most  realistic  EoS of hadron matter at high temperatures and moderate particle number densities. 
First, a long time ago it was found that for temperatures below 170 MeV and moderate  baryonic charge densities 
(below the nuclear saturation density) the  mixture of stable hadrons 
whose interaction is described by  the quantum  second virial coefficients behaves almost  like  a mixture of  ideal gases of  
particles in  which  both the stable hadrons and their 
resonances are included, but  the latter  should have the averaged vacuum values of masses \cite{Raju}. 
As it was demonstrated  in Ref.~\cite{Raju} and recently discussed in Ref.~\cite{Edward2018}, the main physical reason for this kind of  behavior is 
rooted in   an almost complete cancellation between the attractive  and repulsive terms  in  the quantum second virial coefficients. 
Hence, the remaining   deviation from the ideal gas (a weak repulsion) can be modeled  by the  classical hard-core repulsion.

Second,   considering   the HRGM  as the EoS  of hadronic matter one can be sure that  its pressure will never  exceed 
the one of the quark-gluon plasma.
The latter may occur,  if  the hadronic phase  is  modelled  as the mixture of ideal gases \cite{IST3,Satarov10}. 
It is well-known that the number of spin-isospin degeneracies of all known hadrons and their resonances with the masses up to 2.6 GeV is
so large that, if one ignores  the hard-core repulsion between them, at temperatures above 180 MeV their pressure  will be larger than the pressure of the quark-gluon plasma.  
An example of  comparing the HRGM  EoS   with the lattice QCD results  can be seen in  Fig. 8 of Ref. \cite{IST3}.

Third,  an additional  and important reason to consider  the HRGM as the hadronic matter EoS in the vicinity of CFO is a purely  practical one: 
the hard-core repulsion is a contact interaction and, hence, the energy per particle of such an EoS coincides with  the one of the ideal gas, 
even for the case of  quantum  statistics  \cite{ISTQ}. 
Consequently, during the evolution of the system after 
CFO to the kinetic freeze-out  one will not face a hard mathematical problem  \cite{KABkinFO1,KABkinFO2} to somehow 
``transform``  the potential energy of interacting particles into their kinetic energy and into the masses of  particles which appear due to resonance decays. 

These are the main reasons which  allow one to regard  the HRGM as an extension of the statistical bootstrap model \cite{KAB_Ref27}  supplemented  by   the hard-core repulsion, but for  a truncated hadronic mass-volume  spectrum, 
 and which allow one to successfully  apply  it to the description of   hadronic multiplicities measured in 
the central  heavy ion collision experiments. 

Although during  last few years  several  valuable results  were  obtained with the help of  HRGM \cite{MHRGM1,MHRGM2,MHRGM3,MHRGM4,GSA15,GSA16,GSA16b,Signals18,Signals19,KAB_Chatter15}, 
at the moment  the hard-core radii  are well established for the most abundant hadrons,  i.e.  for pions, for the
lightest  K$^\pm$-mesons, for nucleons and for the lightest (anti)$\Lambda$-hyperons.
 Nevertheless,  the} 
 HRGM based on classical virial coefficients is very successful in describing the properties of a hadron gas at CFO temperatures above 50 MeV, hence it is natural  to apply it to the description of multiplicities of atomic nuclei measured 
in A+A collision experiments instead of calculating their quantum virial coefficients. 

However,  even finding the classical  excluded volumes of light (anti)nuclei consisting of $A$ baryons is, in general, a highly nontrivial  task, since there is no well-developed formalism to calculate the cluster integrals of the particles which are clusters themselves.  Due to this reason  the usual Mayer procedure to calculate such cluster integrals cannot be used in the general case. { Furthermore, 
even the classical  excluded volumes of  light (anti)nuclei  with  hadrons are known the next nontrivial task is to rigorously derive the corresponding system of equations which can be used for  the actual fitting of the data. }

Fortunately,  the light  nuclei of $A$ baryons with $A \in [2; 4]$  are roomy clusters,
i.e. their root mean square (rms) radii  ${R}_{\rm rms} = \sqrt{\langle r^2 \rangle }$  are rather large \cite{KAB_Bohr,RecentRms} as one can see from the second column  of  Table 1. This fact allows us  to easily  find out  their classical second virial coefficient with the hadrons if the hard-core radii of all constituents of the considered nuclei  are known. Assuming that the light nuclei of $A$ baryons  can be considered as the quasi-classical particles which slowly move around the common center-of-mass on the distance ${R}_{\rm rms}$,   one can estimate the typical distance between the constituents $L ({R}_{\rm rms})$ in such nuclei.  
For the (anti)deuteron the typical distance between the (anti)nucleons is $L ({R}_{\rm rms}) \simeq 2 ({R}_{\rm rms})$, while to estimate such a distance for  triton, $^3$He,  $^3_{\Lambda}$H and their antiparticles we suppose that   they are  the equilateral triangles.  In this case, ${R}_{\rm rms}$ is the radius of the circle  described around the equilateral triangle and, hence,
the classical  distance between the constituents  inside such nuclei is  $L ({R}_{\rm rms}) \simeq \sqrt{3} R_{\rm rms} \simeq 1.732 R_{\rm rms}$. Similarly, for  the $^4$He nucleus  and its antiparticle,  we assume that the nucleons form an equilateral tetrahedron with the   radius  of   {the} sphere  described  around it being $R_{\rm rms}$. 
Then  the  classical  distance between the constituents of  the $^4$He nucleus is $L ({R}_{\rm rms}) \simeq  4 R_{\rm rms}/\sqrt{6} \simeq  1.633 R_{\rm rms}$.  These simple formulae and the actual estimates of  $L ({R}_{\rm rms})$ for different light nuclei are, respectively,  given in the 3-rd and  4-th columns of Table 1.

\begin{table}[t!]
{\footnotesize 
\begin{center}
\begin{tabular}{|l|c|c|c|}
                                                                                                      \hline
Nucleus             &     R$_{\rm rms}$  & classical distance     &  L         \\ 
		     &  (fm)                        & $L (R_{\rm rms}) $ &  (fm)  \\ \hline
deuteron & $2.1421\pm0.0088$        & $2 R_{\rm rms}$  & 4.280  \\ \hline
triton  &         $1.7591 \pm 0.0363 $  & $\sqrt{3} R_{\rm rms}$ & 3.047   \\ \hline
$^3$He &    $1.9661 \pm 0.0030$     & $\sqrt{3} R_{\rm rms}$    &3.405   \\ \hline
$^4$He  &   $1.6755 \pm 0.0028 $     & $ 4 R_{\rm rms}/\sqrt{6} $ & 2.739   \\ \hline
$^3_{\Lambda}$H &   $4.9$ (Ref. \cite{HtritonR}) & $\sim\sqrt{3} R_{\rm rms}$  & 8.487  \\ \hline
\end{tabular}
\label{table1}
\end{center}
\caption{
The rms radii of light nuclei $R_{\rm rms}$ (2-nd column) taken from \cite{RecentRms}, except for  the $^3_{\Lambda}$H nuclei which is an estimate of Ref. \cite{HtritonR}. The 3-rd  column shows  the relation between the typical distance among the constituents  $L$ and the rms radius $R_{\rm rms}$ of  light nuclei, whereas the 4-th column provides the actual estimates for  $L ({R}_{\rm rms})$.
See text for details.}
}
\end{table}

Comparing  the typical distances  $L ({R}_{\rm rms}^A)$ between the constituents  of $A$ baryons nuclei 
with the sum of  largest  hard-core diameter  of hadrons $0.84$ fm \cite{IST1,IST2} and the hard-core diameter  of baryons $2 R_b = 0.73$ fm \cite{IST1,IST2}, one concludes  that 
it is possible to  freely translate the hadron with the hard-core radius $R_h$ around each of the nucleus constituent, i.e. a baryon of hard-core radius $R_b=0.365$ fm,    without touching  any other constituent  of   this  nucleus \cite{Grinyuk2020}. 
Therefore, the classical  second virial coefficient (excluded volume per particle) of a hadron and a  nucleus of $A$ baryons can be written as
 \begin{equation}\label{Eq1}
 b_{Ah} = b_{hA} =  A \frac{2}{3}\pi (R_b+R_h)^3\,, 
\end{equation}
where $R_b$ is  the hard-core radius of baryons.  

Similarly,  we introduce the classical second virial coefficient (excluded volumes per particle) $b_{h_1 h_2}$ of hadrons of radii $R_{h_1}$ and $R_{h_2}$  as
 	\begin{equation}\label{Eq2}
	b_{h_1 h_2} = b_{h_2 h_1} \equiv \frac{2}{3}\pi(R_{h_1}+R_{h_2})^3 .
	\end{equation}
Now we consider a mixture of hadrons and light nuclei as Boltzmann particles with hard-core interaction. 
Neglecting for a moment the nucleus-nucleus  interaction, i.e. assuming that $b_{A_1 A_2} = 0$,   one can write the
total  excluded volume of such a mixture as 
	\begin{eqnarray} \label{Eq3}
{V}_{excl}^{tot} = \sum\limits_{k \in h_1, A_1} \sum\limits_{ l \in h_2, A_2} N_k  b_{k l} N_l ,
	\end{eqnarray}
where $N_k$ ($N_l$)  is either the number of hadrons of sort $h$ or the number of  nuclei of $A$ baryons.
Note that in the sums in Eq.  (\ref{Eq3}) the antiparticles are considered as the independent sorts of particles.

It is convenient to introduce the additional degeneracy of nuclei of $A$ baryons $g_{kA}$ and explicitly  write  the 
second virial coefficient  (\ref{Eq1})  as
 \begin{eqnarray}\label{Eq4}
&& b_{k h_l} =  g_{kA} \frac{2}{3}\pi (R_k+R_{h_l})^3 = g_{kA} \frac{2}{3}\pi  \times  \nonumber \\
 & & \hspace*{7.5mm}\times (R_k^3 +3R_k^2R_{h_l}+3R_k R_{h_l}^2+R_{h_l}^3),  \\
 \label{Eq5}
{\rm where}~ & &  g_{kA} \equiv A \delta_{kA} + \delta_{kh} , ~ {\rm and} \quad  g_{AA} R_A^n = A R_b^n, 
\end{eqnarray}
where $\delta_{kA}$ and  $\delta_{kh}$ are the Kronecker  $\delta$ symbols. 

Using the fact that the mean number of light nuclei $\langle N_A \rangle$ is very small compared to the mean  number of all other hadrons $\sum_h \langle N_h \rangle$, i.e. $ \langle N_A \rangle \ll \sum_h \langle N_h \rangle$,  for light nuclei we can also write 
 %
 \begin{equation}\label{Eq7}
A \langle N_A \rangle \ll \sum_h \langle N_h \rangle ,
\end{equation}
which allows us to approximate Eq. (\ref{Eq3}) as 
\begin{eqnarray} \label{Eq6}
{V}_{excl}^{tot} & \simeq & \frac{2}{3}\pi  \sum\limits_{k \in h_1, A_1} \sum\limits_{ l \in h_2, A_2} N_k  g_{kA_1}  \times  \nonumber \\
& \times & (R_k^3 +3R_k^2R_{l}+3R_k R_{l}^2+R_{l}^3)
 N_l  g_{l A_2},
\end{eqnarray}
where we substituted  the binomial expression  (\ref{Eq4}) for nucle\-us-hadron interaction and a similar binomial formula  for the  hadron-hadron interaction  
 into Eq.  (\ref{Eq3}). In addition in Eq.  (\ref{Eq6}) the double summation  is extended 
 by adding the second degeneracy  factor $g_{l A_2}$
 to account for the nucleus-nucleus interaction in a symmetric  way which is convenient for further evaluation.   Due to the inequality (\ref{Eq7})
which is valid for light nuclei
the  approximated Eq.  (\ref{Eq6})  is rather accurate for  A+A collisions.  
 
Combining the first  term  in the brackets of  Eq. (\ref{Eq6})  with the last term, and the second term with the third one, 
it is possible to identically  rewrite the total excluded volume   (\ref{Eq6})  in a shorter form 
\begin{eqnarray} \label{Eq8}
{V}_{excl}^{tot} \simeq  \frac{4}{3}\pi \hspace*{-2.2mm} \sum\limits_{k \in h_1, A_1} \sum\limits_{ l \in h_2, A_2}
\hspace*{-2.2mm}  N_k  g_{kA_1}   (R_k^3 +3R_k^2R_{l})  N_l  g_{l A_2},
\end{eqnarray}
which can be used to determine the mean excluded  volume of the system  per particle 
\begin{eqnarray} \label{Eq9}
&&\overline{V}_{excl}  = {V}_{excl}^{tot}  \biggl/  \sum\limits_{ l \in h, A} 
\hspace*{-1.1mm}  N_l   \simeq  {V}_{excl}^{tot}  \biggl/  \sum\limits_{ l \in h, A} 
\hspace*{-1.1mm}  N_l  g_{lA} \simeq  \\
&& \simeq  \hspace*{-2.2mm} \sum\limits_{k \in h_1, A_1} 
\hspace*{-2.2mm}  N_k  g_{kA_1}V_k +  \hspace*{-2.2mm}  \sum\limits_{ k \in h_1, A_1} \hspace*{-2.2mm}   N_k  g_{kA_1} S_k \, \overline{R}  ,
\label{Eq10}
\end{eqnarray}
where we introduced the eigen volume  $V_k = \frac{4}{3}\pi  R_k^3 $ and eigen surface $S_k=  4 \pi  R_k^2$ of the particle of hard-core radius $R_k$ and the mean hard-core radius $\overline{R}$  defined  as 
	\begin{equation}
	\label{Eq11}
	\overline{R} =  \sum\limits_{ k \in h, A} \hspace*{-2.2mm} N_k  g_{kA}  R_k  \biggl/ \sum\limits_{ l \in h, A} 
\hspace*{-1.1mm}  N_l .
	\end{equation}
To obtain Eqs. (\ref{Eq9})-(\ref{Eq11}) we, apparently,  employed the inequality (\ref{Eq7}). 
In the thermodynamic limit
Eqs.  (\ref{Eq10}) and (\ref{Eq11}) enable   us to self-consistently determine  the EoS of the considered mixture within the VdW approximation.

To proceed further,  we assume 
that for an infinite system one can replace all $ N_k$ values in (\ref{Eq11})
by their  statistical mean values $\left\langle N_k\right\rangle $ and write
	\begin{equation}
	\label{Eq12}
	\overline{R} \rightarrow {\sum\limits_{ k \in h, A} \left\langle N_k\right\rangle  g_{kA} R_k}\biggl/ {\sum\limits_{l  \in h, A}\left\langle N_l\right\rangle} .
	\end{equation}
	where $\left\langle N_l\right\rangle$ will be calculated self-consistently using the grand canonical ensemble (GCE) partition function. 
This means that  using $\overline{V}_{excl}$ (\ref{Eq10}) with $\overline{R}$ defined by Eq. (\ref{Eq12}) one can calculate  the GCE partition function regarding $\overline{R}$ as a function of temperature $T$  and chemical potentials 
$\{\mu_k\}$ and afterwards one can find  $\overline{R}$ from the calculated partition.
	
Denoting  the chemical potential  for the $k$-th sort of  particles as $\mu_k$, one can write the GCE partition function as
	\begin{eqnarray}
	&& \hspace*{-4.4mm}Z (T,\left\lbrace \mu_k \right\rbrace, V) \equiv \nonumber
	\\
	&& \hspace*{-4.4mm}\equiv \sum\limits_{ \left\lbrace N_k \right\rbrace }^\infty \hspace*{-0.55mm} \left[ \prod\limits_{k\in h, A}
	\hspace*{-0.55mm} \frac{\left[ \phi_k e^{\frac{\mu_k}{T} }(V - \overline{V}_{excl}) \right]^{N_k}}{N_k!} \right] \theta(V - \overline{V}_{excl}) . \quad
	\label{Eq13}
	\end{eqnarray}

In  Eq.~(\ref{Eq13}) the  thermal density $\phi_k$  of the $k$-th sort of particles
contains  the Breit-Wigner mass attenuation. 
In the Boltzmann approximation  $\phi_k$  can be written as
\begin{eqnarray}\label{Eq14}
\phi_k = g_k  \gamma_S^{|s_k|} \int\limits_{M_k^{Th}}^\infty  \,  &&
\frac{ d m}{N_k (M_k^{Th})} 
\frac{\Gamma_k}{(m-m_{k})^{2}+\Gamma^{2}_{k}/4} \times \nonumber \\
\times
\int  &&\frac{d^3 p}{ (2 \pi \hbar)^3 }   \exp \left[{\textstyle  - \frac{ \sqrt{p^2 + m^2} }{T} }\right] \,,
\end{eqnarray}
where $g_k$ denotes  the degeneracy factor of the $k$-th  sort of particle,
$\gamma_S$ is its  strangeness suppression factor \cite{Rafelski}, 
$|s_k|$ is the number of valence strange quarks and antiquarks in this 
sort of particle, while  the factor 
\begin{equation}\label{Eq15}
\displaystyle {N_k (M_k^{Th})} \equiv \int
\limits_{M_k^{Th}}^\infty \frac{d m \, \Gamma_k}{(m-m_{k})^{2}+
\Gamma^{2}_{k}/4} 
\end{equation}
denotes 
a normalization constant, in which   $M_k^{Th}$ denotes  the decay 
threshold mass of the $k$-th hadronic resonance, while  $\Gamma_k$ denotes its   width.
Clearly, for the stable hadrons and light nuclei the width $\Gamma_k$ should be set to zero,
which leads to the familiar expression for the thermal density
	\begin{equation}\label{Eq16}
	\phi_k = g_k \gamma_S^{|s_k|} \int  \frac{dp^3}{(2\pi\hbar)^3}  \exp \left[{\textstyle  - \frac{ \sqrt{p^2 + m^2_k} }{T} }\right] .
	\end{equation}

We would like  to stress  that the Breit-Wigner ansatz for the mass attenuation is an approximation which is usually valid  for relatively 
narrow resonances only.
However,  the expression for  thermal density  of  unstable particles (\ref{Eq14}) in the spirit of a Beth-Uhlenbeck EoS \cite{Beth:1937zz} 
is valid  for more general mass distributions which may replace this ansatz.
For some dynamical models of hadron structure such as the NJL model, one could separate the resonant part of the interaction which would 
correspond to an unstable hadronic state and can be approximated by a Breit-Wigner ansatz and the residual, repulsive interaction 
\cite{Hufner:1994ma,Wergieluk:2012gd,Blaschke:2013zaa}. This is fortunate if the approach shall be combined with an excluded volume 
model for the short-range repulsion, in order to avoid a possible double counting.  
The generalized Beth-Uhlenbeck EoS can be rigorously derived  for a mixture of  hadron resonances   \cite{KAB_David:16A,KAB_David:16B} 
from a cluster decomposition of the Phi-functional approach \cite{KAB_Phi-approach}, if  the generalized
Phi-functional belongs to  the class of cluster two-loop diagrams  \cite{KAB_Phi-approach2,KAB_Phi-approach3}.

Note that the  Heaviside step function $\theta$ in Eq. (\ref{Eq13}) is very important, since it ensures the absence of negative values of   the available volume $(V - \overline{V}_{excl})$ and provides
the finite number of all particles for finite volume of the system $V$.  However, 
due to its presence,  the evaluation of the GCE partition  function  (\ref{Eq13}) is hard.
To overcome this difficulty one should  make  the Laplace transformation with respect to $V$ to  
the  isobaric partition (for an appropriate review see  \cite{Reuter08}) which is defined as 
	\begin{equation}
	\label{Eq17}
	\hspace*{-2.2mm}
	{\cal Z} (T,\left\lbrace \mu_k \right\rbrace, \lambda) \equiv \int\limits_{0}^{\infty}dV e^{-\lambda V}Z (T,\left\lbrace \mu_k \right\rbrace, V).
	\end{equation}
	Below we show that
the isobaric partition ${\cal Z} (T,\left\lbrace \mu_k \right\rbrace, \lambda)$ can be found  exactly  by changing the integration variable $dV \rightarrow d(V-\overline{V}_{excl})$. However, first of all  it is necessary to define the quantities  $\left\langle N_k \right\rangle $ in the GCE variables. In terms of  the partial $\mu_k$-derivative of the partition  (\ref{Eq13}), one can define $\left\langle N_k \right\rangle$ as follows
	\begin{equation}\label{Eq18}
	\left\langle N_k \right\rangle \equiv T \frac{\partial}{\partial \mu_k} \ln \left[  Z(T,\left\lbrace \mu_l \right\rbrace, V) \right].
	\end{equation}
In terms of definition (\ref{Eq18})  Eq. (\ref{Eq12}) for $\overline{R}$  can be cast  as
	\begin{eqnarray}
	\label{Eq19}
	&&\overline{R} 
	= \frac{\sum\limits_{k \in h, A} g_{kA} R_k \frac{\partial}{\partial \mu_k}\ln  [ Z(T,\left\lbrace \mu_l \right\rbrace, V)] }{
	\sum\limits_{k \in h, A} \frac{\partial}{\partial \mu_k}\ln [ Z(T,\left\lbrace \mu_l \right\rbrace, V)] } .
	\end{eqnarray}
	Changing the variable $dV \rightarrow d(V-\overline{V}_{excl})$ in  Eq. (\ref{Eq17}), one finds
	\begin{eqnarray}
	&&{\cal Z}(T,\left\lbrace \mu_k \right\rbrace, \lambda) = \int\limits_{0}^{\infty}dV' e^{-\lambda V'} \times
\nonumber	\\
\label{Eq20}
	&& \times \sum\limits_{ \left\lbrace N_k \right\rbrace } \prod\limits_{k\in h, A} \frac{1}{N_k!} \left[ \phi_k e^{ \frac{\mu_k}{T}} V'   \right]^{N_k} e^{-\lambda \overline{V}_{excl} } \theta(V') \,.\quad
	\end{eqnarray}
	Substituting   into Eq. (\ref{Eq20}) the expression  (\ref{Eq10}) for $\overline{V}_{excl}$, one gets 
	\begin{eqnarray}\label{Eq21}
	&&\hspace*{-2.2mm}{\cal Z}(T,\left\lbrace \mu_k \right\rbrace, \lambda) = 	\nonumber \\
	&&\hspace*{-2.2mm}= \int\limits_{0}^{\infty}\hspace*{-0.55mm}dV' e^{-\lambda V'} \hspace*{-0.55mm} \sum\limits_{ \left\lbrace N_k \right\rbrace } \prod\limits_{k \in h, A} \frac{ \left[ \phi_k e^{ \frac{\mu_k}{T} -\lambda g_{kA}(V_k +  \overline{R} S_k) }  V'  \right]^{N_k} }{N_k!} =\nonumber \\
	&&\hspace*{-2.2mm}= \int\limits_{0}^{\infty}\hspace*{-0.55mm} dV'  \hspace*{-0.55mm} \exp\left[ V' 
	\hspace*{-0.55mm} \left[ \sum\limits_{k \in h, A} \phi_k e^{\frac{\mu_k}{T} - \lambda g_{kA}(V_k +  \overline{R} S_k) } -\lambda \right]  \right] . ~~~
	\end{eqnarray}
Integration  with respect to variable $dV'$ in Eq. (\ref{Eq21}) can be done easily resulting in
	\begin{equation}\label{Eq22}
	{\cal Z}(T,\left\lbrace \mu_k \right\rbrace, \lambda) = \frac{1}{\lambda - \mathcal{F}(\lambda,T,\left\lbrace \mu_k \right\rbrace)} ,
	\end{equation}			
	where the  function $\mathcal{F}(\lambda,T,\left\lbrace \mu_k \right\rbrace)$ which defines the system pressure in the 
	thermodynamic limit 
	is given by
\begin{equation}\label{Eq23}
	\hspace*{-1.7mm}\mathcal{F}(\lambda,T,\left\lbrace \mu_k \right\rbrace) = \hspace*{-1.4mm} \sum\limits_{k \in h, A}  \hspace*{-0.55mm} \phi_k \hspace*{-0.55mm} \exp\left[ \frac{\mu_k}{T} -  \lambda g_{kA} [V_k +  S_k \overline{R}]  \right] .
\end{equation}		
The GCE partition function (\ref{Eq13}) can be found now  by the inverse Laplace transform 
	\begin{eqnarray}\label{Eq24}
	Z(T,\left\lbrace \mu_k \right\rbrace, V) &=& \frac{1}{2\pi i} \int\limits\limits_{\chi - i\infty}^{\chi + i\infty} \hspace*{-2mm} d\lambda \, e^{\lambda V} \, {\cal Z}(T,\left\lbrace \mu_k \right\rbrace, \lambda) = 
	\nonumber
	\\
	&=& \frac{e^{\lambda^{*} V}}{1-\frac{\partial \mathcal{F}}{\partial \lambda}(\lambda,T,\left\lbrace \mu_k \right\rbrace)}\biggl|_{\lambda = \lambda^{*}} .
	\end{eqnarray}
As usual,  in Eq.  (\ref{Eq24}) the integration contour in the complex $\lambda$-plane is chosen to the right-hand side of the rightmost singularity $\lambda^*$, i.e. $\chi>\lambda^*$ (more details can be found in Ref. \cite{Reuter08}).
Since the number of hadronic states and light nuclei used in the HRGM is finite \cite{IST1,IST2}, then 
the sum in Eq. (\ref{Eq24}) contains the finite number of terms and, hence, as shown in Ref. \cite{Reuter08}, 
the  isobaric partition   (\ref{Eq24}) has only  the simple pole at $\lambda=\lambda^*$. The latter is  a solution of the equation
\begin{equation}\label{Eq25}
	\lambda^{*} = \mathcal{F}(\lambda^{*},T,\left\lbrace \mu_k \right\rbrace). 
\end{equation}
In the thermodynamic limit $V \rightarrow \infty$
from Eq. (\ref{Eq24}) one finds the system pressure as $p \equiv T \lambda^*$ , since  in this limit the GCE partition behaves  as $Z(T,\{\mu_k\},V \rightarrow \infty) \sim \exp(pV/T )$ \cite{Huang}.

Using Eq. (\ref{Eq25}) one can write for the pressure
	\begin{equation}\label{Eq26}
	p = T \sum\limits_{k \in h, A} \phi_k \exp\left[ \frac{\mu_k- p g_{kA} [V_k  +  \overline{R} S_k] }{T} \right] ,
	\end{equation}
which should be supplemented by 
	the equation for the mean hard-core radius $\overline{R}$. Using Eq. (\ref{Eq24}) one can rewrite Eq. (\ref{Eq19})
	as follows
	\begin{eqnarray}\label{Eq27}
	&& \overline{R} = \frac{\sum\limits_{k \in h, A}  g_{kA} R_k \frac{\partial}{\partial \mu_k} \left[ \lambda^{*} V  - \ln (1-\frac{\partial \mathcal{F}}{\partial \lambda^{*}} )\right]  }{\sum\limits_{k \in h, A} \frac{\partial}{\partial \mu_k} \left[ \lambda^{*} V  - \ln (1-\frac{\partial \mathcal{F}}{\partial \lambda^{*}} )\right] } \,.
	\end{eqnarray}
In the thermodynamic limit  $V \rightarrow \infty$  the terms  $ \ln (1-\frac{\partial \mathcal{F}}{\partial \lambda^{*}} )$ 
in Eq.  (\ref{Eq27}) are small  compared to the term $\lambda^* V$.  Hence,  
finding the partial derivatives $\frac{\partial \lambda^{*}}{\partial \mu_k}$  from Eq. (\ref{Eq25}), 
in the limit  $V \rightarrow \infty$
 one can rewrite Eq. (\ref{Eq27}) as
	\begin{eqnarray}\label{Eq28}
	&& \overline{R}  = 
	\frac{\sum\limits_{k  \in h, A}  g_{kA} R_k \phi_k \exp\left[ \frac{\mu_k - p g_{kA}[V_k + S_k \overline{R} ]}{T} \right] }{\sum\limits_{k  \in h, A}  \phi_k \exp\left[ \frac{\mu_k - p g_{kA}[V_k + S_k \overline{R} ]}{T} \right]}
.
	\end{eqnarray}
With the help of equation (\ref{Eq26}) for pressure  it is convenient to cast the last result in terms of the induced surface tension 
 (IST) coefficient \cite{IST1}
	\begin{eqnarray}\label{Eq29}
	\Sigma &\equiv&  p \overline{R} = \nonumber \\
	&=& T \hspace*{-1.1mm} \sum\limits_{k  \in h, A} \hspace*{-1.1mm}  g_{kA} R_k \phi_k \exp\left[ \frac{\mu_k - p g_{kA} V_k - \Sigma g_{kA} S_k}{T} \right] . \quad 
	\end{eqnarray}
Rewriting   equation for pressure similarly, one gets
	\begin{eqnarray}
	\label{Eq30}
	p & = &  \sum\limits_{k \in h, A} p_k =  \nonumber \\
	& = & T \sum\limits_{k \in h, A} \phi_k \exp\left[ \frac{\mu_k - p g_{kA} V_k - \Sigma g_{kA} S_k}{T} \right]\,,
	\end{eqnarray}
where the partial pressures $\{ p_k\}$ of each sort of particles are  introduced for convenience.

The  system of Eqs. (\ref{Eq29}) and (\ref{Eq30}) for the IST coefficient $\Sigma$ and pressure $p$, respectively,  defines the EoS 
of the mixture of hadrons and light nuclei within 
 the VdW  approximation.   Note that in contrast to the heuristic derivation of such a system suggested in \cite{Grinyuk2020}
the present derivation of the system (\ref{Eq29}) and (\ref{Eq30}) is rigorous and well-controlled.
The applicability range of the VdW approximation is, unfortunately, rather narrow  and,  therefore,
its usage at the packing fractions $\eta = \sum\limits_{k \in h, A} g_{kA} V_k \rho_k$ 
(here $\rho_k = \frac{\partial p}{\partial \mu_k}$  is the particle number density) 
above 0.12-0.15 may lead to problems with causality \cite{IST2,IST3,Satarov_Cs} 
(a typical example of acausal HRGM  can be found in Ref. \cite{Vovchenko-ALICE}, see also its critique in Refs. \cite{IST2,IST3}). 

Fortunately,   the applicability range of the  system (\ref{Eq29}) and (\ref{Eq30}) can be extended to 
higher values of packing fractions  in a simple way.  
The main  idea of  the  IST approach \cite{IST1,IST2,IST3,QSTAT2019,Nazar2019} is that at high pressures the mean radius $\overline{R}$ in Eqs. (\ref{Eq26}),
(\ref{Eq8})  and (\ref{Eq29})  should be suppressed  stronger than it is provided by the VdW approximation. 
Then for increasing pressure   the mean radius $\overline{R}$  should gradually vanish leading to a reduction of the 
effective excluded volume of particle of $k$-th  sort  which is defined as
\begin{eqnarray}\label{Eq31}
V_k^{eff} &=& g_{kA}\frac{(V_k p + S_k \Sigma )}{p} \rightarrow \nonumber \\
&\rightarrow & \left\{ \begin{array}{lc}
g_{kA}(V_k  + S_k \overline{R} ) , & {\rm for}~~ \frac{ \max[V_h] p}{T} \ll  1,  \\
 & \\
g_{kA} V_k , & {\rm for}~~ \frac{ \max[V_h] p}{T} \gg  1.  \\
\end{array} \right. 
\end{eqnarray}
In other words,  a gradual vanishing of the mean hard-core radius $\overline{R}$ should provide a slow transformation of the VdW (excluded volume)  approximation, which is valid at low packing fractions $\eta \le 0.1$, into the eigen volume approximation,
which is valid at high packing fractions $\eta > 0.5$ \cite{Huang,Simple-Liquids}. 
As suggested in  Ref. \cite{IST1} and verified in Refs. \cite{IST2,IST3,QSTAT2019,Nazar2019} such an additional suppression of $\overline{R}$ can be obtained   by replacing  the term $\Sigma S_k$ on the right hand side of Eq.  (\ref{Eq29}) as 
	\begin{equation}
	\label{Eq32}
	\Sigma S_k \rightarrow \Sigma S_k \alpha_k, \qquad {\rm where} \quad \alpha_k >1
	\,,
	\end{equation}
where the auxiliary parameters $\alpha_k$ should be chosen   in such a way that they describe the higher virial coefficients. 
Under this generalization  Eq. (\ref{Eq29}) becomes 
	\begin{eqnarray}\label{Eq33}
	\hspace*{-0.5mm}\Sigma  &=& \hspace*{-1.1mm}\sum\limits_{k  \in h, A} \Sigma_k =  \nonumber \\
	&=&  T \hspace*{-1.1mm}\sum\limits_{k  \in h, A} \hspace*{-1.1mm}g_{kA} R_k \phi_k \exp \hspace*{-1.1mm}\left[\frac{\mu_k - g_{kA}(p V_k +\alpha_k \Sigma S_k)}{T} \right] ,\quad 
	\end{eqnarray}
where $ \Sigma_k$ denotes the surface tension coefficient of $k$-th sort of particles.
In this way one can account not only for the second virial coefficients, but also for the higher order virial coefficients as demonstrated  for  systems  with 
single-component hard-core repulsion in Refs. \cite{IST2,IST3,QSTAT2019}  and for two-component mixtures  studied recently  in \cite{Nazar2019}. 

The reason to chose all the parameters $\alpha_k$ as $\alpha_k > 1$ becomes  apparent after analyzing  the effective excluded volume (\ref{Eq31}). 
Indeed, substituting Eq.  (\ref{Eq33}) into Eq. (\ref{Eq31}) one finds for hadrons 
\begin{eqnarray}\label{Eq34}
V_k^{eff} & \equiv&  V_k +  S_k \frac{ \sum\limits_{l \in h, A} p_l R_l e^{-(\alpha_l-1)S_l \Sigma/T } }
{ \sum\limits_{n \in h, A} p_n }  . \quad 
\end{eqnarray}
This equation  shows that in the limit of  low packing fractions, i.e. for $\Sigma \max[S_h] /T \ll 1$, each exponential in Eq. (\ref{Eq34}) 
is  $\exp \left[- \frac{(\alpha_l-1)S_l \Sigma}{T} \right] \simeq 1$ and, hence, one recovers the upper Eq. (\ref{Eq31}).  
However,  it is easy to show that for high packing fractions an opposite inequality $\frac{\Sigma {S_k}}{T} \gg 1$ is valid  for  any $S_k >  0$.  
In this case the condition $\alpha_k > 1$  provides the vanishing of  the mean radius  $\overline{R} \equiv \frac{\Sigma}{p}$ and, hence,  
in this  limit the  effective excluded volume of each  particle  approaches its  eigen volume,  $V_k^{eff} \rightarrow V_k  $.  
Thus, at high packing fractions Eq. (\ref{Eq34})  leads to the lower Eq. (\ref{Eq31}).

The system  (\ref{Eq30}), (\ref{Eq33}) is a generalization of the IST EoS derived in Ref. \cite{Nazar2019} for the classical hard spheres onto the multicomponent  mixture  of hard spheres (hadrons)  and  roomy  classical clusters which are the light nuclei of $A$ baryons.  
As one can see from the derivation above such a generalization is not straightforward and contains some nontrivial  steps.  
In particular, the inequality (\ref{Eq7}) played a crucial role in simplifying our derivation of the mean excluded volume per particle. 
Furthermore,  the fact that the thermal density (\ref{Eq14})  of considered particles may, in principle,  include the finite width 
opens an entirely new possibility to apply the present approach to the treatment of other roomy exotic clusters which have even larger width, 
than the hypertriton  $^3_{\Lambda}$H like, e.g., $^4$Li and $^4$H \cite{Ropke:2020peo,Bazak:2020wjn}.

In fact, the system  (\ref{Eq30}), (\ref{Eq33})
can be  generalized further in the  spirit of Refs. \cite{QSTAT2019,Nazar2019} in order to extend it to very high packing fractions $\eta \simeq 0.45-0.5$  by  introducing into treatment   the induced curvature  tension.  

In Refs. \cite{IST2,IST3} it is shown that 
even with   a single   parameter $\alpha_k = const = \alpha =1.245$ Eqs. 
 (\ref{Eq30}), (\ref{Eq33})   for the classical hard spheres 
allows  one  to go beyond the VdW  approximation, whereas in Ref. \cite{Nazar2019} one can find several examples on how two  auxiliary parameters enables us  to go far beyond the VdW  approximation for two component classical  systems. 
However, an extension of the  system  (\ref{Eq30}), (\ref{Eq33})  onto the  quantum mechanical  treatment of  
light (anti)nuclei in the spirit of Ref.  \cite{QSTAT2019} still remains a challenge for theoreticians. 

\section{Analysis of light nuclei multiplicities measured in A+A collisions} \label{Sect3}

The system  (\ref{Eq30}), (\ref{Eq33})  is  the IST EoS with classical excluded volumes of (light) nuclei and, hence, hereafter it is called IST  EoS in order to distinguish it from another treatment of hard-core repulsion developed in 
\cite{KAB_Ref3,Grinyuk2020}.  Although an  approach of Refs. \cite{KAB_Ref3,Grinyuk2020} is approximative,
nevertheless, we consider it as a complementary one to the IST model.  It is based on the idea to introduce 
 the equivalent hard-core radius  $R^{eq}_{Ah}$
of a pair $Ah$  by equating the  excluded volume $\frac{2}{3}\pi (R_{Ah}^{eq})^3$ with the equivalent hard-core radius $R^{eq}_{Ah}$ to the  actual excluded volume of  such a pair $b_{Ah}$ given by Eq. (\ref{Eq1}). Then we get
the equivalent hard-core radius as \cite{Grinyuk2020}
 \begin{equation}\label{Eq35}
 R_{Ah}^{eq} =  A^\frac{1}{3}(R_b+ R_h)\, .
\end{equation}
From the expression for $ R_{Ah}^{eq}$ one can  determine the effective hard-core radius of a nucleus 
in a hadronic medium dominated by pions 
 \begin{equation}\label{Eq36}
R_A \simeq R_{A\pi}^{eq} - R_\pi \simeq  A^\frac{1}{3}R_b + (A^\frac{1}{3}-1) R_\pi \simeq  A^\frac{1}{3}R_b\, .
\end{equation}
It is necessary to stress that 
this approximation is well  justified for the A+A of high energies by the fact that pions are  the most abundant particles.
In particular, this  is the case for the  LHC  and {the} highest RHIC 
collision energies.  
The term $(A^\frac{1}{3}-1) R_\pi$ in Eq.  (\ref{Eq36}) 
is a small correction 
to the effective hard-core radius of nuclei $R_A  \simeq  A^\frac{1}{3}R_b$, 
 since  the hard-core radius of pions $R_\pi \simeq 0.15$ fm  \cite{IST1,IST2} is essentially smaller than the one of baryons $R_b =0.365$ fm and the hard-core radii of   kaons  $R_K=0.395$ fm and other mesons  $R_m=0.42$ fm.
 Therefore, for low values of baryonic chemical potential (roughly for $\mu_B < T$)  the pions are the least suppressed by the  hard-core repulsion  and, consequently,  for 
any $A \le 4$ the correction $(A^\frac{1}{3}-1) R_\pi \le 0.088$ fm  
in Eq.  (\ref{Eq36}) 
can be safely  neglected for the pion-dominated hadronic medium.   
\begin{figure}[ht]
	\centerline{\includegraphics[width=0.85\columnwidth]{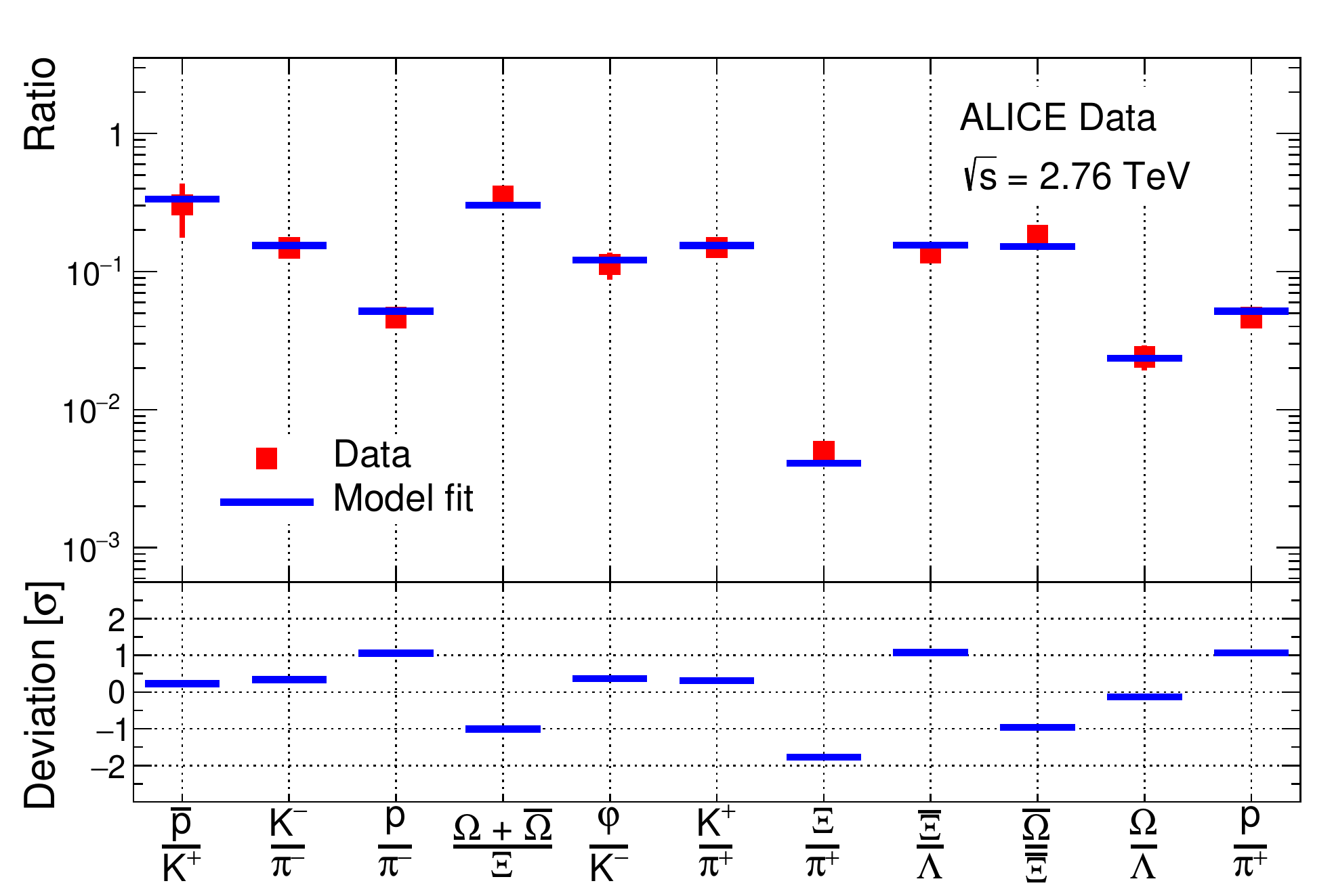}}
	\vspace*{-1mm}
	\centerline{~\includegraphics[width=0.85\columnwidth]{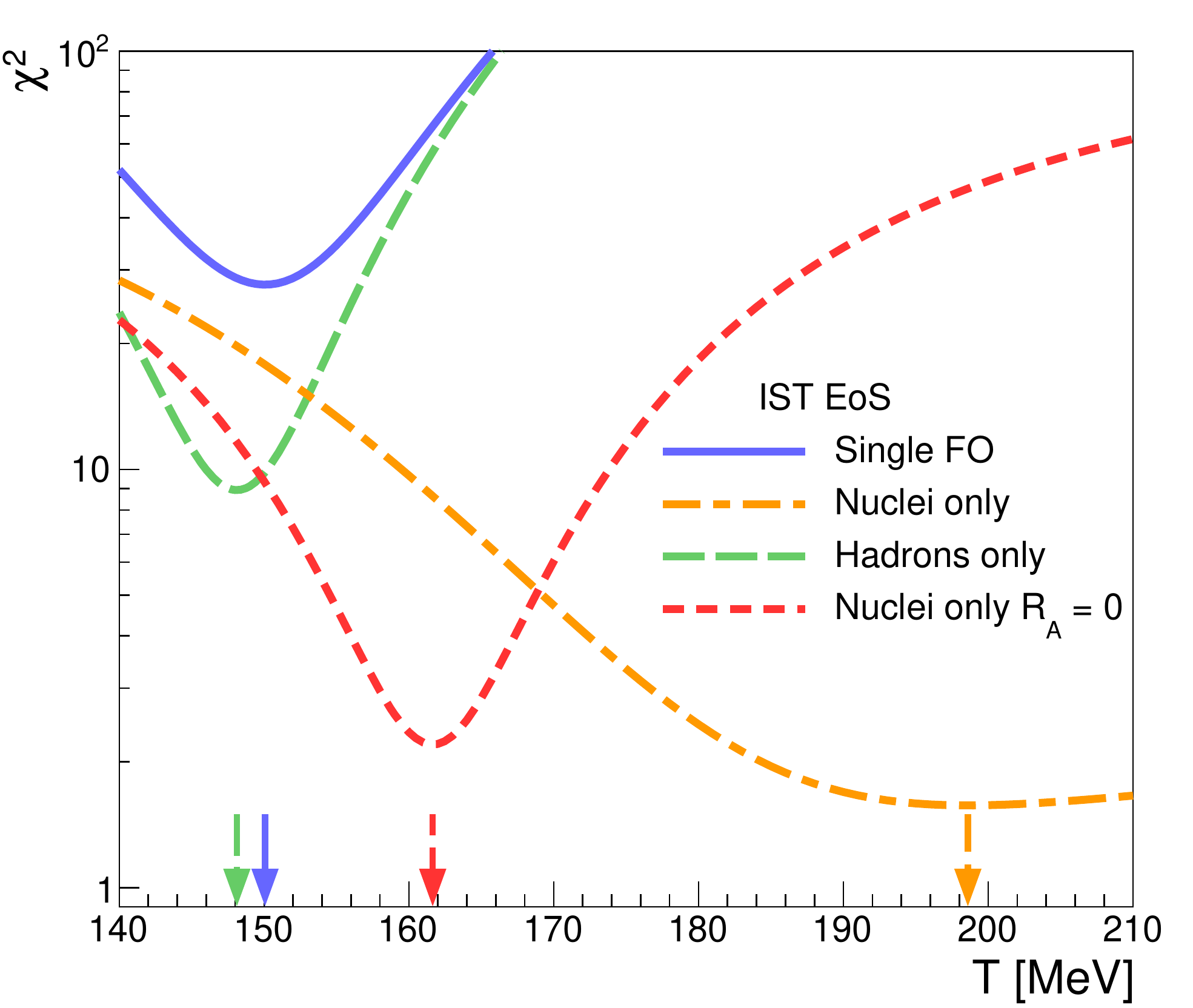}}
\vspace*{-1mm}
	\centerline{~\includegraphics[width=0.85\columnwidth]{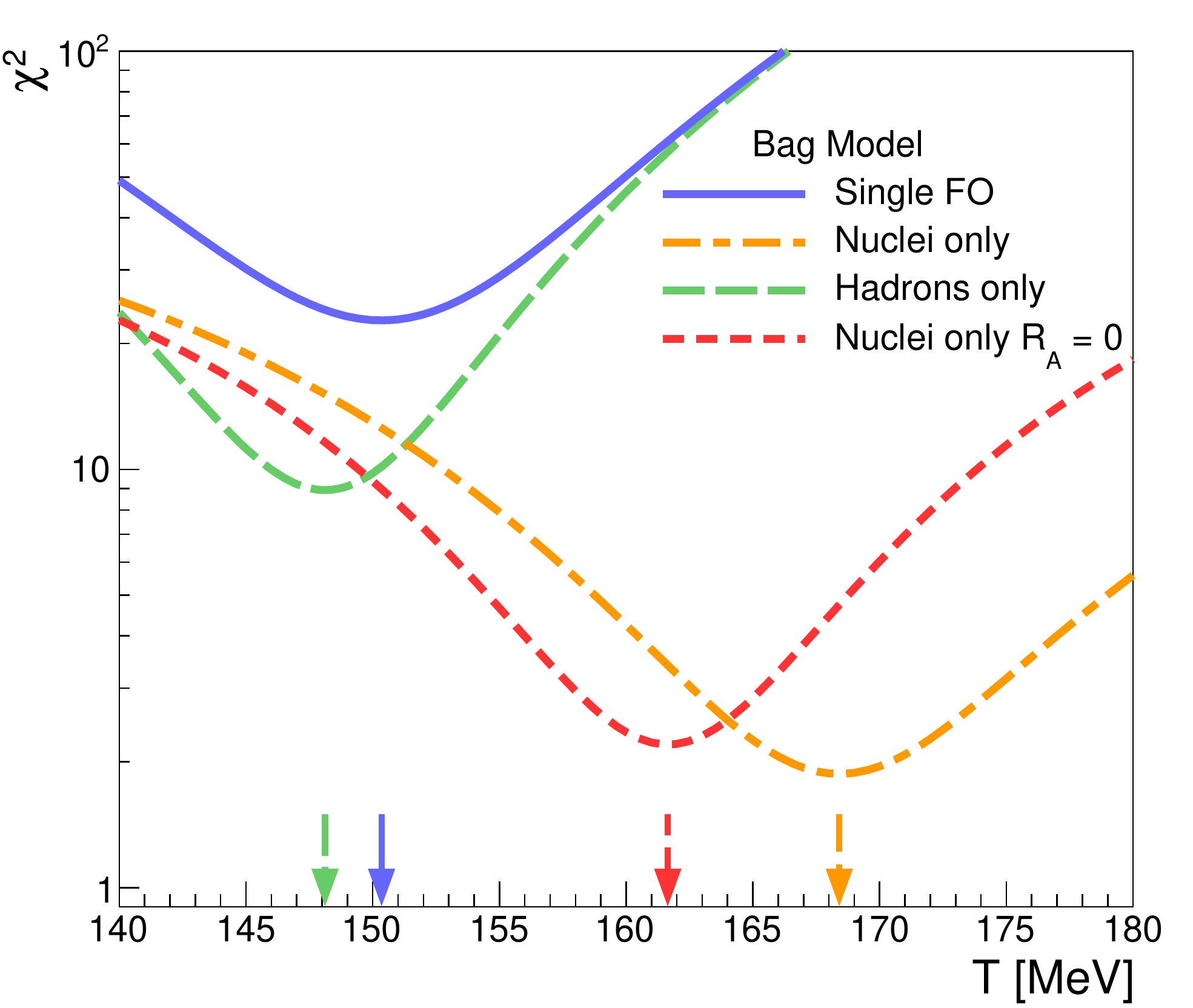}}
\vspace*{-2mm}
	\caption{\small {\bf Upper panel:} Ratios of hadronic yields measured 
	at $\sqrt{s_{NN}} =2.76$ TeV (symbols) vs.  the results of  IST EoS  (\ref{Eq30}), (\ref{Eq33})  (bars, for more details see the text).
	The  CFO temperatures $T_A=T_h = 150.1 \pm 1.9$ MeV  are for the singe CFO IST EoS. 
	Insertion shows the deviation of theory from data in the units of experimental error. 
	{\bf Middle panel:} Temperature 
	dependence of $\chi^2_{tot}$,  $\chi^2_h$ and  $\chi^2_A$ for the 
	IST EoS.
	   {\bf Lower panel:} Same as in the middle  panel, but for the BMR EoS.
	   }
	\label{KAB_Fig1}
\end{figure}

The hard-core radius of light (anti)nuclei (\ref{Eq36})  is similar to the expression of the Bag Model radius (BMR) \cite{MITBagM} of large bags of quark-gluon plasma and, hence, hereafter  this  model is called the BMR EoS. 
Despite the fact that it is an approximative approach, it is, however, simpler because with the help of the hard-core radius (\ref{Eq36})  the IST  EoS allows one to treat the nuclei and hadrons exactly on  the same footing. Moreover, 
{\it a simultaneous use of  IST and BMR  approaches allows  us to introduce a new strategy to locate the CFO of light (anti)nuclei.} 
Since in the pion-dominated hadronic medium  the BMR approach should give the same results as the IST, we have to 
search for the region of parameters at which both approaches provide a similar quality of the data description. 
%

%
For the BMR approach the system   (\ref{Eq30}), (\ref{Eq33})  should be slightly modified. Then 
formally  considering 
the light (anti)nuclei as the primed sorts of hadrons $h_A^\prime$ (with $A=2, 3, 4$), we can write
	\begin{eqnarray}\label{Eq37}
	\hspace*{-2.2mm}p & = &  \sum\limits_{k \in h, h_A^\prime} p_k = \hspace*{-0.25mm}
	 T  \hspace*{-1.25mm} \sum\limits_{k \in h, h_A^\prime} \phi_k \exp\left[ \frac{\mu_k - p V_k - \Sigma  S_k}{T} \right],~ \\
	\label{Eq38}
	\hspace*{-2.mm}\Sigma  &=& \hspace*{-0.15mm}\sum\limits_{k  \in h, h_A^\prime} \hspace*{-1.25mm} \Sigma_k =  
	\nonumber \\
	&=&  
	T \hspace*{-01.1mm}\sum\limits_{k  \in h, h_A^\prime} \hspace*{-01.1mm} R_k \phi_k \exp \hspace*{-0.5mm}\left[\frac{\mu_k - p V_k -\alpha_k \Sigma S_k}{T} \right] ,\quad 
	\end{eqnarray}
where the hard-core radii of primed hadrons are given by Eq. (\ref{Eq36}). 

The partial  values $p_k$ and $\Sigma_k$ entering the system (\ref{Eq30}), (\ref{Eq33}) (or (\ref{Eq37}), (\ref{Eq38})) allow  one  to write the particle number density of $k$-th sort of particle in a simple  way
\begin{equation}\label{Eq39}
\rho_k \equiv \frac{\partial  p}{\partial \mu_k} = \frac{1}{T} \cdot \frac{p_k \, a_{22} 
- \Sigma_k \, a_{12}}{a_{11}\, a_{22} - a_{12}\, a_{21} } \,.
\end{equation}
For the system (\ref{Eq37}), (\ref{Eq38}) the coefficients $a_{kl}$  are given by  \cite{IST2}
\begin{eqnarray}\label{Eq40}
&& a_{11} =  1 + \frac{4}{3} \pi  \hspace*{-01.1mm} \sum\limits_{k  \in h, h_A^\prime} \hspace*{-01.1mm}  R_k^3 \frac{p_k}{T} ,  ~~
a_{12}  =  4 \pi \hspace*{-01.1mm} \sum\limits_{k  \in h, h_A^\prime} \hspace*{-01.1mm} R_k^2 \frac{p_k}{T}  , \\
\label{Eq41}
&&a_{21} = \frac{4}{3} \pi \hspace*{-01.1mm}\sum\limits_{k  \in h, h_A^\prime}\hspace*{-01.1mm} R_k^3\frac{\Sigma_k}{T}  , ~~
a_{22} = 1 + 4 \pi \hspace*{-01.1mm}\sum\limits_{k  \in h, h_A^\prime}\hspace*{-01.1mm}  R_k^2 \alpha_k \frac{\Sigma_k}{T} , \qquad
\end{eqnarray}
while in order to calculate the particle number densities for the system (\ref{Eq30}), (\ref{Eq33})  
in Eqs. (\ref{Eq40}) and (\ref{Eq41}) one has to make 
the following replacements 
\begin{equation}\label{Eq42}
R_{h_A^\prime}^2 \rightarrow A R_b^2, \qquad R_{h_A^\prime}^3 \rightarrow A R_b^3 , 
\end{equation}
for the  powers of hard-core radius of $A$ baryons nucleus. 
After finding all  partial values $\{ p_k\}$ and  $\{ \Sigma_k\}$,
from the expressions  (\ref{Eq39})-(\ref{Eq41})
one can determine the  thermal yield $N_k^{th} = V \rho_k$  of the 
 $k$-th sort of particles.
For hadrons, however,  one has also to add the contribution coming from the decays of resonances.
For the known  branching ratios  $Br_{l\rightarrow k}$ of hadronic  decays $l\rightarrow k$  one can 
write  the total yield of $k$-th sort of  hadrons as follows
\begin{eqnarray}
\label{Eq43}
N^{tot}_k = V\biggl( \rho_k+\sum_{l\neq k}\rho_l\, Br_{l\rightarrow k} \biggr) \,,
\end{eqnarray}
where $V$ is the CFO volume.   Since all  details of the fitting process   are  well presented in the original works \cite{IST2,IST3}, here we discuss the most important issues only. 

To analyze the ALICE data \cite{KAB_Ref1a,KAB_Ref1b,KAB_Ref1c} we use the setup of Ref. \cite{IST2}, while for 
the analysis of the STAR data that of Ref. \cite{IST3}.   
The main difference in fitting the hadrons and the $A$-baryon nuclei is that for hadrons we use the ratios 
\begin{eqnarray}
\label{Eq44}
{\cal R}_{kl}^{theo}  = \frac{\rho_k+\sum_{n\neq k}\rho_n\, Br_{n \rightarrow k}}{\rho_l+\sum_{n\neq k}\rho_n\, Br_{n\rightarrow l}}\,,
\end{eqnarray}
of yields of hadrons of sorts $k$ and $l$.  On contrary for the $A$ baryon nuclei we employ the yields. 
Hence  the total $\chi^2_{tot}(V)$ used in the present work is 
\begin{eqnarray}\label{Eq44}
&&\hspace*{-1.2mm}\chi^2_{tot}(V) = \chi^2_{h} + \chi^2_{A} (V)  = \nonumber \\
&&\hspace*{-1.2mm} = \hspace*{-1.55mm}\sum_{ {k \neq l} \in h} \hspace*{-0.55mm} \left[  \frac{{\cal R}_{kl}^{theo} - {\cal R}_{kl}^{exp}}{\delta {\cal R}_{kl}^{exp}}\right]^2  \hspace*{-0.55mm}+  \hspace*{-0.55mm} \sum_A \left[  \frac{\rho_A(T) V - N^{exp}_A}{\delta N^{exp}_A}\right]^2  . \quad \quad
\end{eqnarray}
Here $\chi^2_{h}$ and $\chi^2_{A}$ denote, respectively,  the  mean deviation squared for hadrons and (anti)nuclei. 
Note that  $\chi^2_{tot}(V)$  is a function of  the CFO volume $V$.  This is an important difference from our previous analyses of \cite{MHRGM1,MHRGM2,MHRGM3,Sagun14,IST2,IST3}, which, as we will show below, allows us to elucidate 
the new details on the CFO of light (anti)nuclei. Since now on we also consider a single value $\alpha_k = 1.25$\cite{IST2,IST3}.

\begin{figure}[t]
	
	\centerline{\includegraphics[width=0.95\columnwidth]{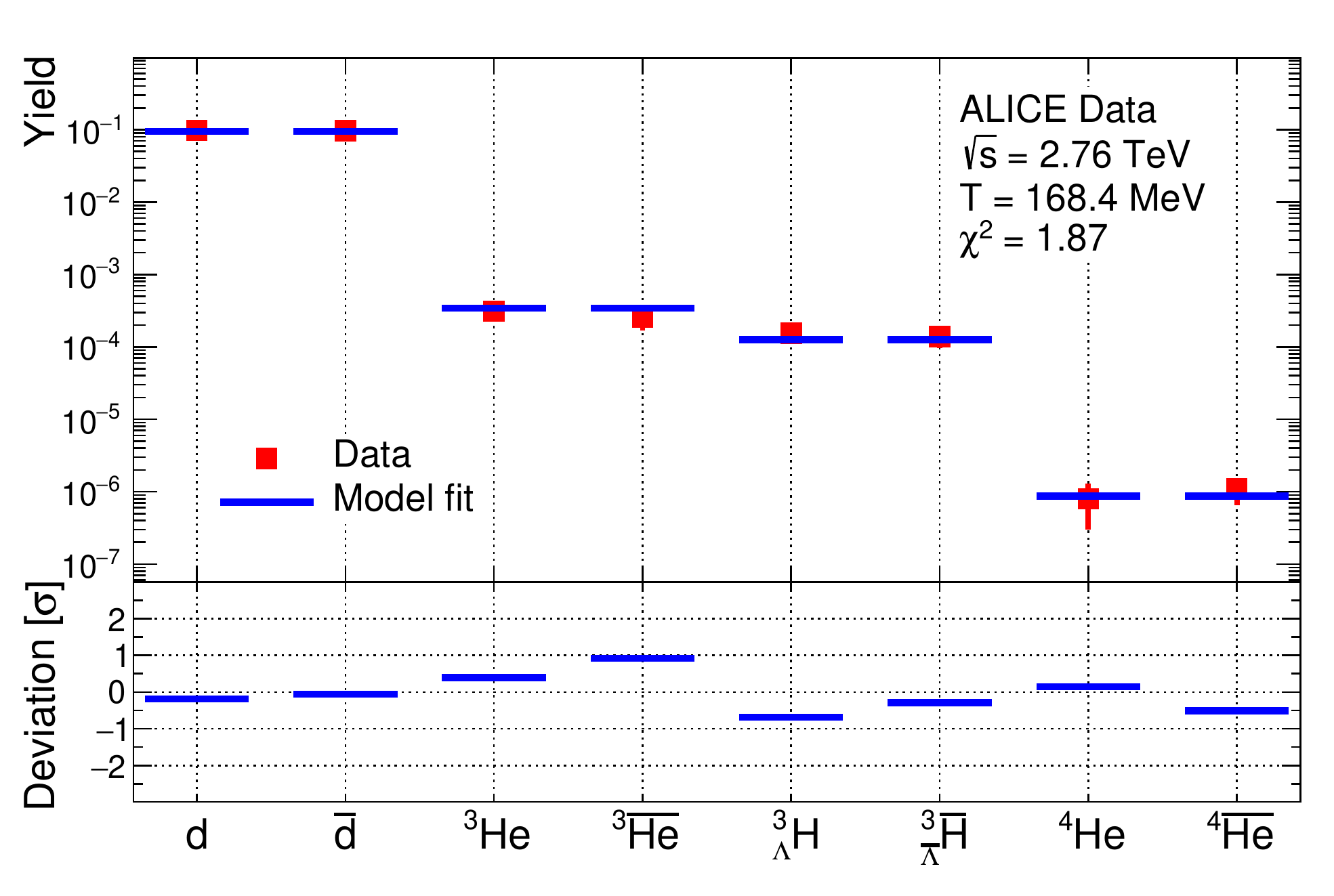}}
	\vspace*{-1mm}
	\centerline{\includegraphics[width=0.95\columnwidth]{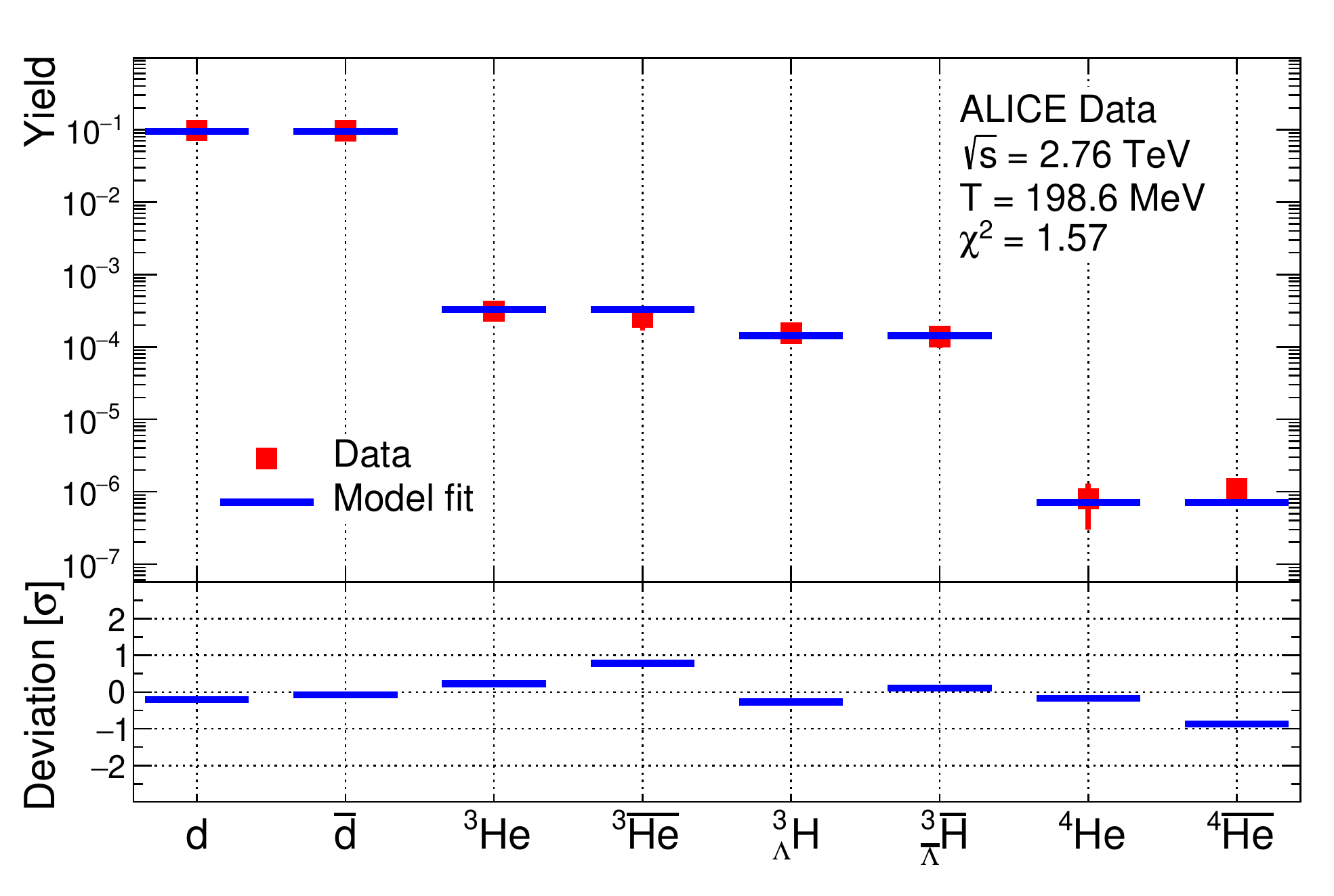}}
\vspace*{-2mm}
	\caption{The yields of nuclear clusters measured at $\sqrt{s_{NN}} =2.76$ TeV vs. theoretical description  in the scenario of separate CFO of light (anti)nuclei.  
	{\bf Upper panel:} The $\min\chi^2_A (V)$ corresponds to the BMR EoS (third row in Table 2).
	   {\bf Lower panel:} Same as in the upper panel, but for the IST EoS 
(forth row in Table 2).
}
	\label{KAB_Fig2}
\end{figure}

	\begin{table*}[t!]
		\centering
		\begin{tabular}[t]{lcccc}
		\toprule
			Description            & $T_h, $~ MeV      & $T_A, $~ MeV      & $V_A, $~ fm$^3$    & $\chi^2/dof$  \\ 
		\midrule
			Single CFO, BMR   & $150.35 \pm 1.91$ & $150.35 \pm 1.91$ & $11241 \pm 2016$ & $1.336$  \\ 

			Single CFO, IST   & $150.06 \pm 1.94$ & $150.06 \pm 1.94$ & $13357 \pm 2277$ & $1.627$  \\ 

			Separate CFO, BMR & $148.12 \pm 2.03$ & $168.41 \pm 5.60$ & $2997 \pm 1164$  & $0.675$  \\ 

			Separate CFO, IST  & $148.12 \pm 2.03$ & $198.59 \pm 30.47$& $1544 \pm 1027$  & $0.656$  \\ 
\bottomrule
		\end{tabular}
		\caption{The results obtained by the advanced HRGM  for the fit of ALICE data measured at $\sqrt{s} = 2.76$ TeV. The CFO temperature of hadrons is $T_h$, the CFO temperature of light (anti)nuclei is $T_A$, while their CFO volume is $V_A$. The last column gives the fit quality. } 
	\end{table*}

First, we apply  the  single CFO model to the ALICE data description. 
{ The hadronic data were taken from Refs.  \cite{Abelev:2013vea,Abelev:2013zaa,Abelev:2013xaa,Abelev:2015prc}.}
In the HRGM  it is  traditionally  assumed that  the CFO  occurs for  all particles simultaneously.
The principal results are given in Table 2 and  Figs.   \ref{KAB_Fig1} and  \ref{KAB_Fig2}.
 To get these results we calculated the $\chi^2_{tot}$ using 2 fitting parameters, i.e. the  CFO temperature and the CFO volume of nuclei $V=V_A$, for 11 hadronic ratios and 8 yields of  light (anti)nuclei. All the chemical potentials  are set to zero, while $\gamma_s =1$  is fixed according to Refs.\cite{IST2,IST3}. 
 The hard-core radii of  hadrons 
  are taken from our previous works \cite{IST2,IST3} (are listed above). These values provide an excellent description of 
 hadron yield ratios from AGS to LHC energies. 

As one can see from Fig. \ref{KAB_Fig1}  and  from Table 2 the quality of ALICE data description  obtained for the single CFO scenario is similar for the IST and BMR EoS. Moreover, the corresponding CFO temperatures are very similar, since the $\chi^2_{tot}$ is completely defined by the hadronic contribution to $\chi^2_{tot}$. 
Although the obtained overall description is satisfactory with $\chi^2_{tot}/ dof |_{IST} \simeq 1.627$ and
$\chi^2_{tot}/dof |_{BMR}\simeq 1.336$,
there are two surprising features in this scenario. First,  $\chi^2_{tot}/ dof |_{IST}$ found by the advanced model  is somewhat larger than $\chi^2_{tot}/dof |_{BMR}$.
 Second, the CFO volumes of light (anti)nuclei are essentially larger than the CFO volume 
$V_h = \frac{N_{\pi^+}^{exp}}{\rho_{\pi^+} +\sum_{l\neq k}\rho_l\, Br_{l\rightarrow {\pi^+}}} \simeq 8165 \pm 600$ fm$^3$ of hadrons
found from the experimental multiplicity of positive pions  $N_{\pi^+}^{exp}$. 
{ This is best seen, when applying the new strategy to determine the common CFO volume for nuclei  which is found to be $V_A \in [11100; 13260] $ fm$^3$. Comparing $V_A$ with  $V_h$ one finds that $\min V_A \simeq 11100$ fm$^3$ is sizably larger than $\max V_h \simeq 8765$ fm$^3$. This means that at the same CFO temperature  the emission volume of hadrons and nuclei are rather different, i.e. the nuclei are freezing out in a much larger volume which means  that there is no common hyper-surface of CFO.}
 In our opinion,  both of these features evidence for  the internal inconsistency of the single CFO scenario at ALICE  energy of collisions. 
Therefore, following the original idea of Ref. \cite{KAB_Ref3}, we verify the hypothesis of separate CFO of light (anti)nuclei.  

From Fig. \ref{KAB_Fig1} one can see that at high CFO temperatures the quantity $\chi^2_{A} (V_A(T_A))$ 
has a deep minimum not only for the IST and BMR EoS, but even for  the vanishing 
size  of nuclei. 
In other words, {the} existence of   a minimum of $\chi^2_{A}$ at high temperatures   is a generic feature of the advanced versions of HRGM. 
In the scenario of separate CFO of nuclei, there are three fitting parameters, namely the  CFO temperatures of hadrons $T_h$
and nuclei $T_A$, and the CFO volume of  nuclei $V_A$. As one can see from Table 2 and from Figs. \ref{KAB_Fig1} and \ref{KAB_Fig2} one can see that   the hypothesis of  separate CFO of nuclei provides an excellent fit 
with $\chi^2_{tot}/dof|_{IST}  \simeq 0.656$ and $\chi^2_{tot}/dof|_{BMR}  \simeq 0.675$.
Thus, 
compared to the single CFO scenario  the value of $\chi^2_{tot}/dof$ 
in this case   decreased by  50\%.

\begin{figure}[t]	
	\centerline{\includegraphics[width=0.95\columnwidth]{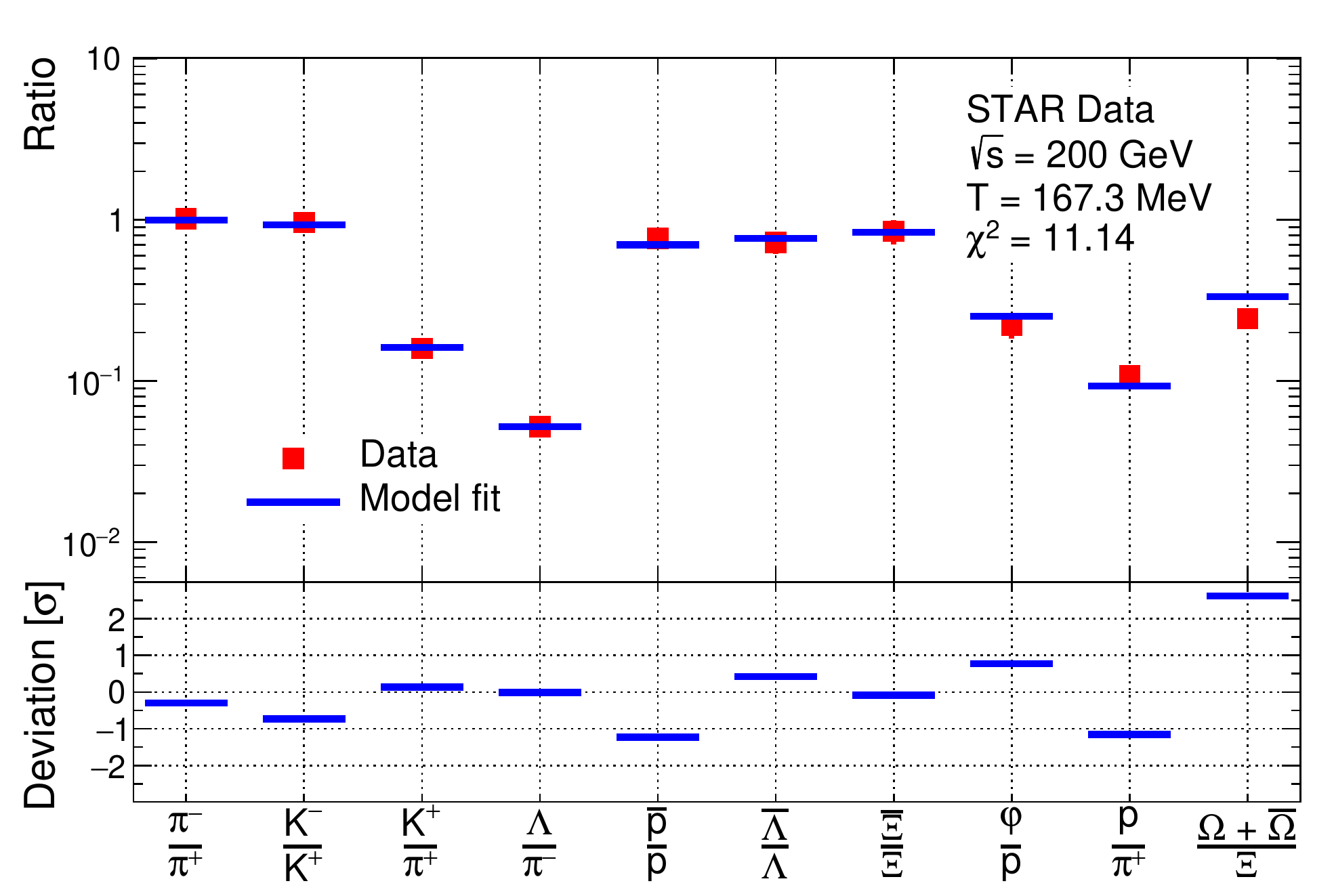}}
	\vspace*{-1mm}
	\centerline{\includegraphics[width=0.95\columnwidth]{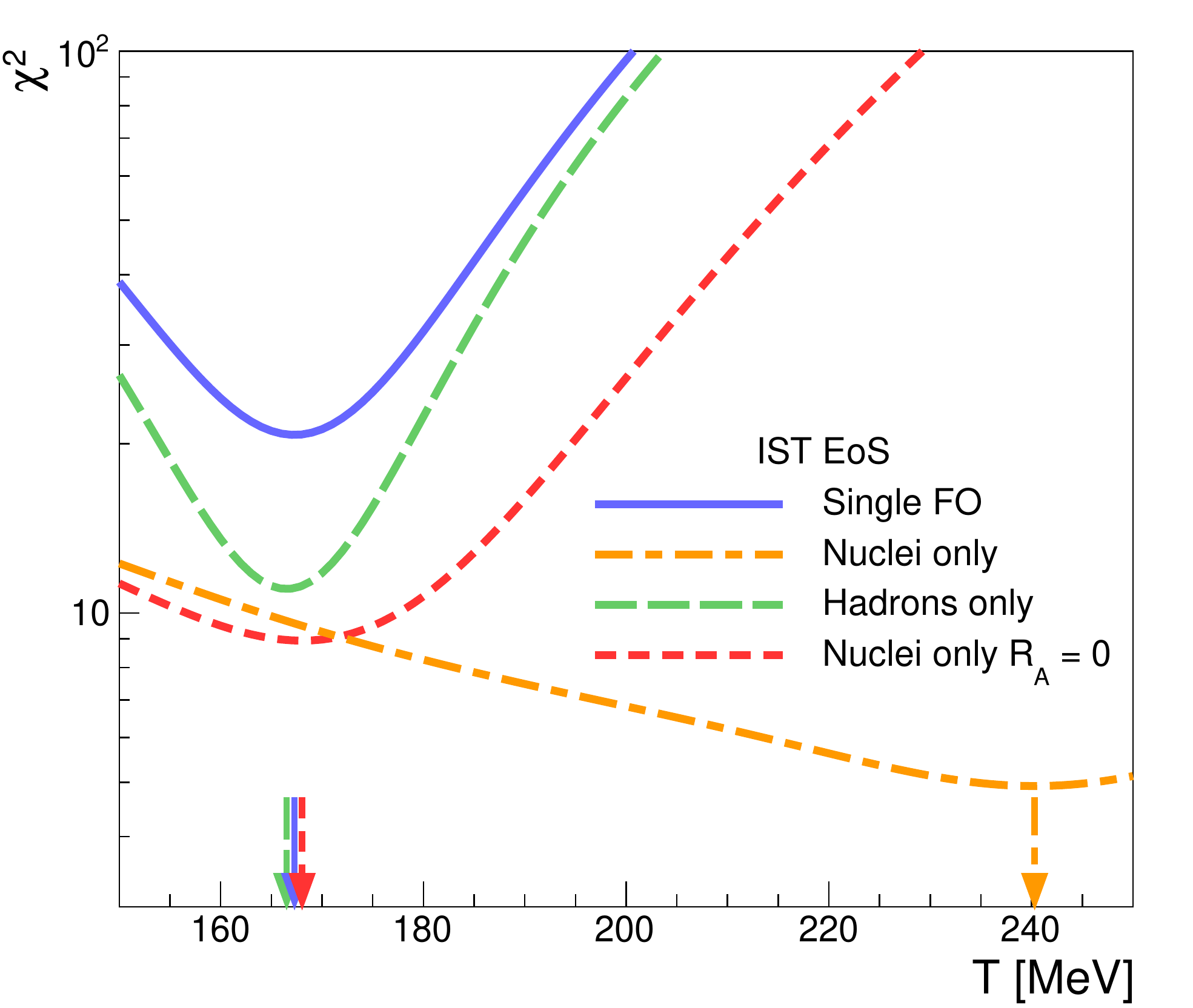}}
\vspace*{-2mm}
	\caption{ {\bf Upper panel:} Ratios of hadronic yields measured
	at $\sqrt{s_{NN}} =200$ GeV (symbols) vs.  the results of  IST EoS (bars) (\ref{Eq30}), (\ref{Eq33})  (for more details see the text).
	The  CFO temperatures  $T_A=T_h = 167.28 \pm 3.93$ MeV  are  given for the singe CFO IST EoS. 
	Insertion shows the deviation of theory from data in the units of experimental error. 
	{\bf Lower panel:}  Temperature 
	dependence of $\chi^2_{tot}$,  $\chi^2_h$ and  $\chi^2_A$ for the 
	IST EoS.
}
	\label{KAB_Fig3}
\end{figure}

However, the minimum of $\chi^2_A |_{IST}$ is located at essentially larger CFO temperature than  the one found for the 
minimum of $\chi^2_A |_{BMR}$. 
Moreover, the found $T_A$ for IST EoS is so large, that one can doubt the existence of hadrons and nuclei at this CFO temperature. 
Fortunately, with the help of the new strategy introduced in the preceding section one can resolve this problem easily. 
Indeed,  similar results for the description of light (anti)nuclei 
by the IST and BMR EoS can be achieved in the vicinity of the common  CFO temperature  $T_A^{com}$ defined by the equality
\begin{eqnarray}\label{Eq46}
&&\chi^2_A (V(T_A^{com}))  |_{IST} = \chi^2_A (V(T_A^{com}))  |_{BMR} ~\Rightarrow  \\
&& \Rightarrow ~ T_A^{com}= 175.1^{+2.3}_{-3.9}~{\rm MeV} ,
\label{Eq47}
\end{eqnarray}
where the common CFO temperature still corresponds to a very accurate description of the ALICE data for light (anti)nuclei 
$\chi^2_A (V(T_A^{com}))  |_{IST} \simeq 3.2$ and $V_A^{com}= V_A(T_A^{com}) \simeq 2660^{+1010}_{-1160}$ fm$^3$. The upper and lower deviations from $T_A^{com}= 175.1$ MeV
in Eq. (\ref{Eq47})  were found numerically by increasing  the value of $\chi^2_A (V(T_A^{com}))  |_{IST} \simeq 3.2$  on  1$\sigma$.
Note that for $T_A^{com}= 175.1$ MeV the total value of   $\chi^2_{tot} / dof = 12.123/16 \simeq 0.758$ is rather small, i.e. it  still corresponds to a  highly  accurate description of the ALICE data. 
Furthermore, the found  range of  $T_A^{com}$  is consistent with the values of  CFO temperature  found for the RHIC  collision energy  \cite{IST2,KAB_Jean,KAB_Chatter15,KAB_pbm17} and it is a few MeV above the upper estimate
for the   
cross-over temperature $T_{co} \simeq  147-170$ MeV  predicted by the lattice formulation of QCD at vanishing value of the  baryonic chemical potential \cite{KAB_lqcd1,KAB_lqcd2}. Therefore, we are confident that the HRGM is applicable at 
these values of the common CFO temperature.

It is necessary to mention that the above numbers  differ slightly from our preliminary results of a similar analysis reported in Ref. \cite{Grinyuk2020}. The main difference is that in the present work we use the non-vanishing width for all hadronic resonances, while in Ref. \cite{Grinyuk2020} the solutions of systems  (\ref{Eq29}), (\ref{Eq30}) and  (\ref{Eq37}), (\ref{Eq38}) were found for zero width of all hadronic resonances in order to fasten the fit process. However, the value of 
$T_A^{com}= 175.1$ MeV found here and the result   $T_A^{com}= 174.6$ MeV found in \cite{Grinyuk2020} are practically the same, whereas  its uncertainty determined here is a couple of MeV larger than in Ref.  \cite{Grinyuk2020}.

\begin{figure}[t]
	
	\centerline{\includegraphics[width=0.95\columnwidth]{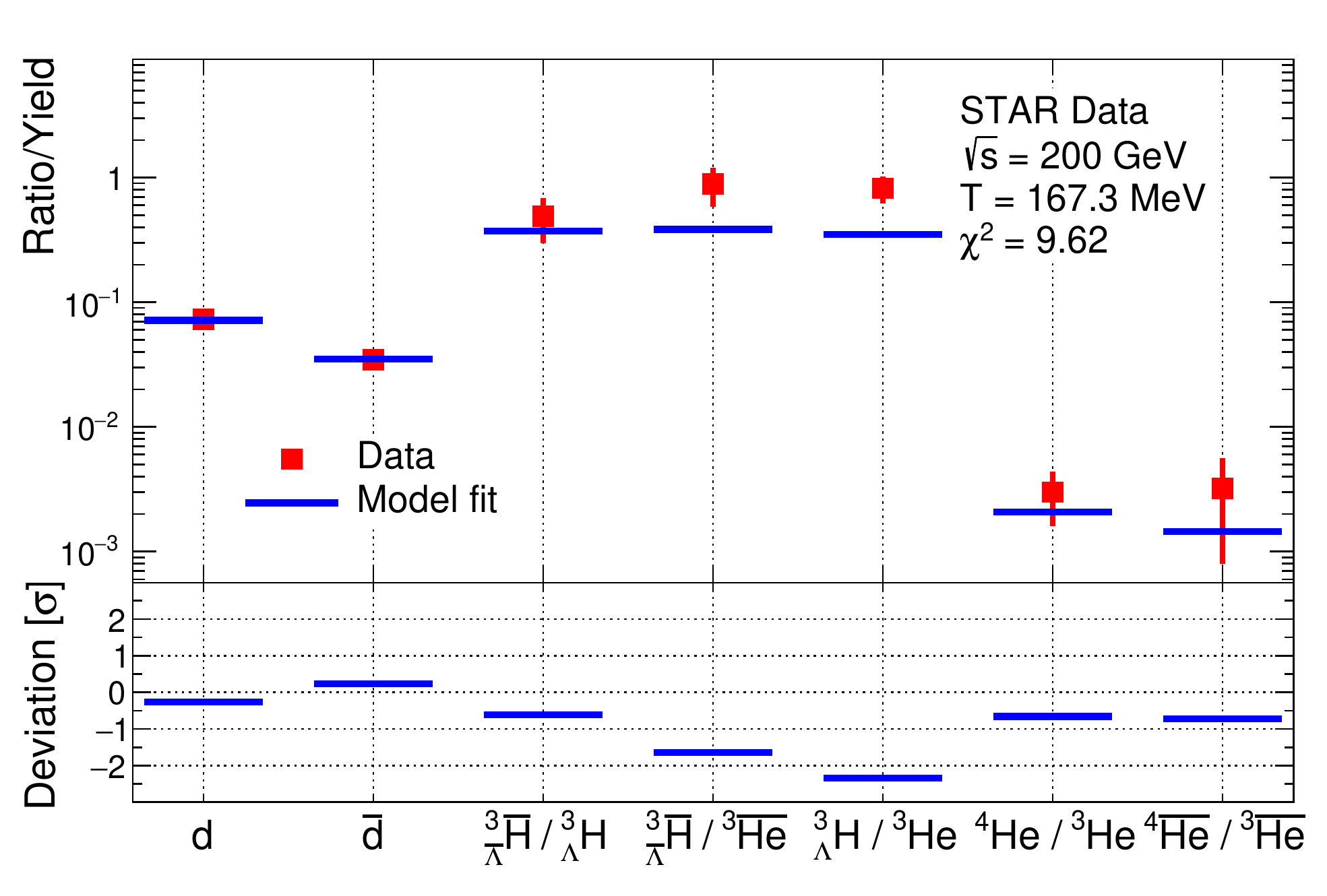}}
	\vspace*{-1mm}
	\centerline{\includegraphics[width=0.95\columnwidth]{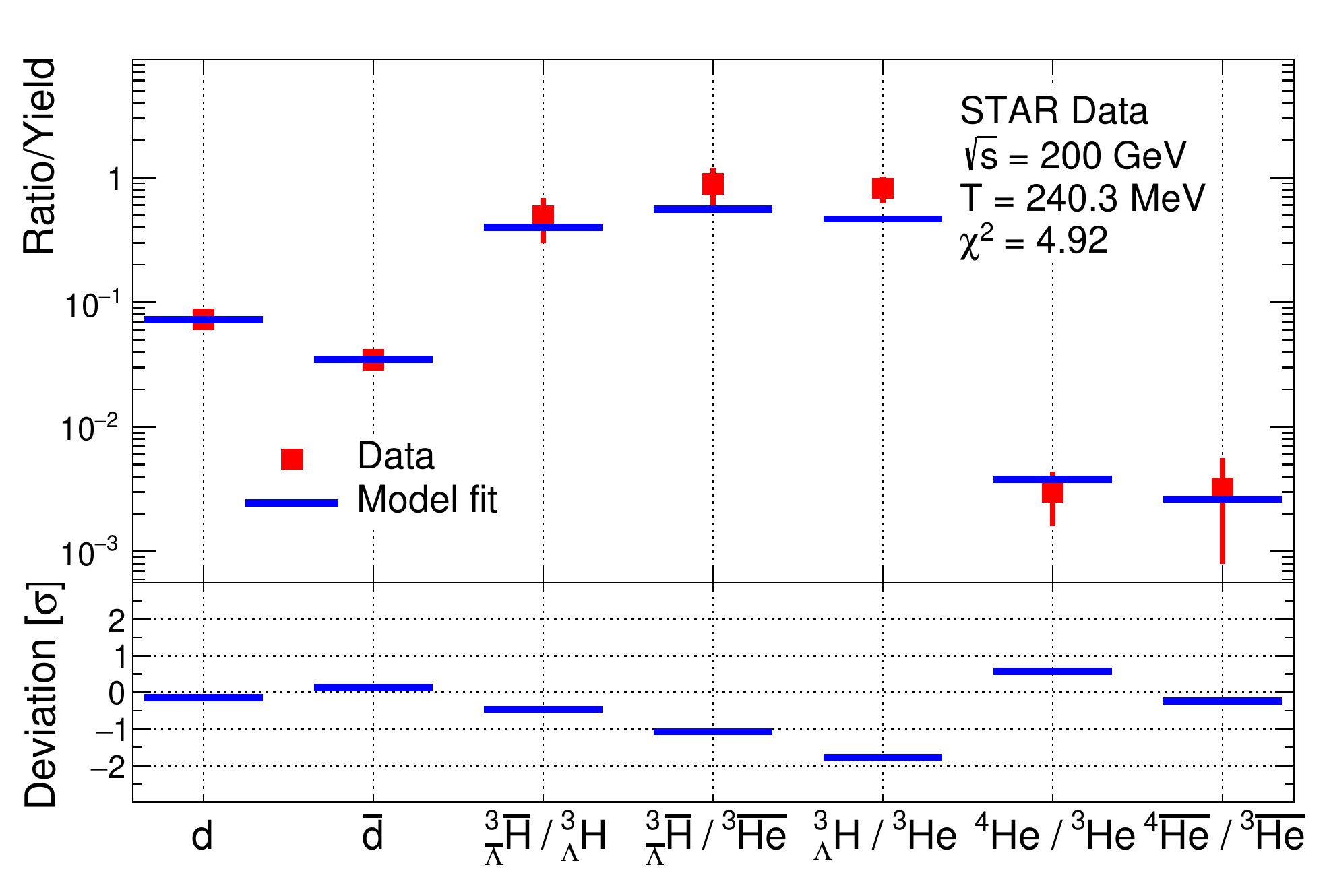}}
\vspace*{-2mm}
	\caption{ {\bf Upper panel:} Yield of (anti)deuteron and ratios of yields of light (anti)nuclei measured 
			at $\sqrt{s_{NN}} =200$ GeV (symbols) vs.  the results of  IST EoS (bars)  (\ref{Eq30}), (\ref{Eq33})  (for more details see the text).
			The  temperatures $T_A=T_h = 167.28 \pm 3.93$ MeV  are for the singe CFO with  IST EoS. 
			Insertion shows the deviation of theory from data in the units of experimental error. 
			{\bf Lower panel:} Same as in the upper   panel, but for the separate CFO  with IST EoS for  
			$T_A = 240.29 \pm 21.38$ MeV.}
	\label{KAB_Fig4}
\end{figure}

The results of a similar  analysis  of  the STAR data measured at  $\sqrt{s} = 200$ GeV are presented in Table 3 and  Figs. \ref{KAB_Fig3} and \ref{KAB_Fig4}. 
The STAR data consist of 10 hadronic ratios { that are  taken from Refs.  \cite{KABstar:200a,KABstar:200b,KABstar:200c} and  are  shown}  in the upper panel of  Fig. \ref{KAB_Fig3}, yields of (anti)deuterons \cite{STARA3},   and 5 light (anti)nuclei yield ratios  \cite{STARA1,STARA2}. 
For the single CFO scenario,  we have 3 fitting parameters, i.e. CFO temperature $T_h = T_A$, CFO baryonic chemical potential $\mu_B^A=\mu_B^h$ and the CFO volume of nuclei  $V_A$.  
The strange chemical potential  and the one of  third projection of isospin are set to zero for simplicity, while $\gamma_s =1$ according to Ref. \cite{IST3}. 

The results obtained for the  hadronic ratios are depicted in the upper panel of Fig.  \ref{KAB_Fig3}, while the CFO temperature scan of  $\chi^2_{tot} (V(T_A=T_h))  |_{IST}$
is shown in the lower panel of this figure.  As one can see from Fig.  \ref{KAB_Fig3} all hadronic ratios, except  the ratio $\frac{\Omega + \bar \Omega}{\Xi} $,  are  well reproduced by the IST EoS. From Table 3 one can see that the CFO temperature
$T_h$, the CFO baryonic chemical potential $\mu_B^h$  and $\chi^2_{tot} / dof$
obtained for the IST and BMR EoS are practically the same.  But the most striking result is that for the single CFO scenario,
the value of common CFO volume
$V_A^{com} = 1898.5  \pm 157.5$ fm$^3$ (see Table 3) is  only 30 percent  smaller compared to the corresponding value
$V_A^{com} \simeq 2660^{+1010}_{-1160}$ fm$^3$ found above for the ALICE energy.
{\it We believe this is a remarkable finding, since the collision energy of  the ALICE data  is about 14 times larger than the one of the STAR data.}
At the same time for this scenario,   the CFO volume of hadrons $V_h = 2808 \pm 253$ fm$^3$ found via the density of positive pions is slightly  larger.

	\begin{table*}[t!]
		\centering
		\begin{tabular}[t]{lcccccc}
			\toprule
			Description            & $T_h, $~ MeV        & $T_A,$ ~ MeV        &$\mu_{B}^{h},~$ ~ MeV&$\mu_{B}^{A}, $~ MeV & $V_A,$ ~ fm$^3$       & $\chi^2/dof$  \\ 
			\midrule
			Single CFO, BMR   & $ 167.16 \pm 3.87 $ & $ 167.16 \pm 3.87 $ & $ 29.99 \pm 3.25 $ & $ 29.99 \pm 3.25 $ & $ 1692 \pm 364 $ & $ 1.429 $  \\ 
			
			Single CFO, IST     & $ 167.28 \pm 3.93 $ & $ 167.28 \pm 3.93 $ & $ 30.05 \pm 3.26 $ & $ 30.05 \pm 3.26 $ & $ 2155 \pm 411 $ & $ 1.482 $  \\ 
			
			Separate CFO, BMR & $ 166.51 \pm 4.07 $ & $ 178.62 \pm 14.63 $ & $ 28.84 \pm 5.37 $ & $ 32.63 \pm 4.94 $ & $ 979 \pm 605 $ & $ 1.607 $  \\ 
			
			Separate CFO, IST   & $ 166.51 \pm 4.07 $ & $ 240.29 \pm 21.38 $ & $ 28.84 \pm 5.37 $ & $ 44.08 \pm 6.81 $ & $ 545 \pm 537 $ & $ 1.330 $  \\ 
			\bottomrule
		\end{tabular}
		\caption{The results obtained by the advanced HRGM  for the fit of STAR data measured at $\sqrt{s} = 200$ GeV. The CFO temperature of hadrons (nuclei) is $T_h$ ($T_A$), the CFO baryonic chemical potential  of hadrons (nuclei) is 
			$\mu_B^h$
			($\mu_B^A$), while the CFO volume of nuclei  is $V_A$. The last column gives the fit quality.} 
	\end{table*}

For the scenario of separate CFO of  light (anti)nuclei,  the CFO temperature of nuclei is substantially higher  than the one of hadrons as one can see from Table 3.  Although the higher  value of $T_A$ provides a better description of the light (anti)nuclei yields as it is  seen  from Fig.  \ref{KAB_Fig4}, the value of  
$\chi^2_{tot} (V(T_A^{com})) / dof  |_{IST}  = \chi^2_{tot} (V(T_A^{com})) / dof  |_{BMR}  \simeq 1.61$  with $T_A^{com} \simeq 180 \pm 11.25$ MeV  is about 
10 percent larger than for the scenario of a single CFO. 
The reason is that for the scenario of separate CFO there are two additional parameters, namely the temperature of nuclei $T_A$ and their baryonic chemical potential $\mu_B^A$ were fitted in this case. 
As a result, the number of degrees of freedom in this case is $17-5 = 12$.  
Therefore, in contrast to the ALICE data,  the STAR data do not demonstrate any preference for the separate CFO of light (anti)nuclei.  
Furthermore, in our opinion, there is no reason to expect that CFO of light (anti)nuclei can occur at the CFO temperatures above 175-180 MeV, since the hadronic description at those CFO temperatures is rather problematic according to the contemporary lattice version of QCD   \cite{KAB_lqcd1,KAB_lqcd2}.
 { 
The recent  lattice QCD result for the continuum-extrapolated chiral susceptibility at vanishing values of baryonic chemical potential define with high accuracy the 
pseudo-critical transition temperature $T_{pc}=156.5\pm 1.5$ MeV \cite{Bazavov:2018mes} from its peak position. 
However, it also demonstrated that this peak has a full-width at half maximum of about 40 MeV.  
Therefore we conclude that  the STAR data which favor the single CFO scenario with $T_A^{com}  |_{STAR} \simeq  167.22 $ MeV and 
$V(T_A^{com})  |_{STAR} \simeq  1898.5 $ fm$^3$ (see Table 3) 
is not in contradiction with the lattice QCD result of Ref. \cite{Bazavov:2018mes}.
 }
 
It is necessary to point out that the lower quality of the description of STAR data is generated by the ratios  $_\Lambda^3 \overline{H}/ ^3 \overline{He}$ and $_\Lambda^3 {H}/ ^3 {He}$ (see Fig. \ref{KAB_Fig4}).
This is an old puzzle \cite{PBM2010,Rapp2018} which still awaits for its solution.  
It is clear, however, that an increase of the CFO temperature above 175-180 MeV is not a viable solution and hence one has to look for another explanation.

The most intriguing question of this work is: how can one interpret the common CFO volume of light (anti)nuclei $V_A^{com}$? 
Of course, at the present stage of research, this question cannot answered with confidence, since neither the mechanism of light nuclei production nor the mechanism of their thermalization are well established. 
However,  our educated guess is that  the thermal production of light (anti)nuclei  is  most naturally  caused by  the 
{ 
hadronization of quark-gluon bags   \cite{KAB_Ref3,Chapline:1978kg} 
} 
formed in A+A  collisions  which  have  the Hagedorn mass spectrum  \cite{KAB_Ref27}.  
Since the Hagedorn mass spectrum is a perfect thermostat and a perfect particle reservoir \cite{Thermostat1} any particle or cluster emitted by  the bags with  
such a mass spectrum  will be produced in full chemical and thermal equilibrium with the emitting bag.  
Such a hypothesis is not only able, in principle,  to  explain  the fact that the light nuclear clusters appear in full chemical and thermal  equilibrium, but also  the 
fact that the  CFO  temperatures extracted here from the ALICE and STAR data coincide.
{ 
Moreover, in a recent preprint \cite{Gallmeister:2020fiv} based on a simplified transport model which, nevertheless, accurately takes into account  the microscopic  reactions between hadrons and heavy resonances with the Hagedorn mass spectrum it is shown that  such an approach is able to reasonably well reproduce the ALICE data on hadronic and light nuclei multiplicities \cite{KAB_Ref1a,KAB_Ref1b,KAB_Ref1c} with the common CFO temperature of hadrons and nuclei $T_{com}=149$ MeV for the Hagedorn temperature $T_H = 167$ MeV. 
Note that this value of $T_{com}$ almost coincides with the result  $T_h=T_A =150.35\pm1.91$ MeV obtained within a single CFO scenario  for  the  ALICE data (see Table 2). 
Unfortunately,  the authors of Ref. \cite{Gallmeister:2020fiv}  have not analyzed the case of different CFO temperatures for hadrons and light nuclei, but the results obtained in the present work clearly demonstrate that this scenario is more favorable. 
}

Therefore, in accordance with our hypothesis,   the found values of $V_A^{com}$ can be considered as the total  volume of 
all  quark-gluon bags  from which the light (anti)nuclei are produced. 
If  this is the case, we can  predict  that the entropy of quark-gluon  bags produced at the collision energies $\sqrt{s_{NN}} =2.76$ TeV and  
$\sqrt{s_{NN}} =200$ GeV are related to each other as
\begin{eqnarray}\label{Eq48}
\frac{S_{ALICE}}{S_{STAR}} \simeq \frac{V_A^{com} \cdot (T_A^{com})^3 |_{ALICE}}{V_A^{com}  \cdot  (T_A^{com})^3 |_{STAR}} \simeq  1.609 .
\end{eqnarray}
Since during  the hydrodynamic expansion of a perfect fluid  the entropy is approximately conserved, the ratio of entropies at CFO of nuclei should be equal to the ratio of initial entropies formed at the moment of thermalization of  quark-gluon bags. 
Therefore, the relation between initial entropies (\ref{Eq48}) can be either verified by the hydrodynamic simulations or, alternatively, it can help
to fix the value of  initial  energy density which is used in the integrated Hydro Kinetic Model \cite{IHKm}.

In fact,  the  common CFO volumes obtained for two different energies of collision allow us to determine the number of emitting  sources of nuclei.
From  the ratio of two common CFO volumes  $ \frac{V(T_A^{com})  |_{ALICE}}{V(T_A^{com})  |_{STAR}} = 1.4015 \simeq \frac{7}{5} = \frac{14}{10}$ one can find the radius of emitting source $R_A^{source} \simeq 4.49$ fm for the number of 7 sources for the ALICE data and 5 sources for STAR data, or $R_A^{source} \simeq 3.566$ fm for the number of 14 sources for the ALICE data  and 10 sources for the STAR data. 
Of course, it may be just a coincidence, but the radius of emitting source $R_A^{source} \simeq 3.566$ fm is just 0.25 percent smaller than  the coalescence model parameter  $\delta r  = 3.575$ fm used  in Ref. \cite{Coalesc1} to model the formation process of light (anti)nuclei. 
Therefore, according to our hypothesis  that the light (anti)nuclei are produced from the quark-gluon bags with Hagedorn mass spectrum   at
the moment of their hadronization the radius of the emitting source $R_A^{source} \simeq 3.566$ fm is, most likely, the radius of  such bags. 
However, an additional verification of the found emitting source radius is necessary. 

\section{Conclusions} \label{Conclusions}
In this work,  we suggested and exploited  an entirely new strategy to elucidate the CFO parameters of light (anti)nuc\-lei produced 
in A+A collisions of high energy, in which the medium of secondary  hadrons is dominated by pions. 
This strategy is based on two different approaches to model the hard-core repulsion between  light nuclei and hadrons. 
The first approach is based on an approximate treatment of  equivalent hard-core radius of  roomy nuclear clusters and pions.
The second approach is rigorously derived here using  a self-consistent treatment of classical excluded  volumes of light nuclei  and hadrons.  In other words, here we generalized the induced surface tension concept to the mixtures of hadrons of different hard-core  radii and light (anti)nuclei of different sizes and masses,  and derived the corresponding equation of state. 

Since in the pion-dominated  hadronic medium both approaches should give the same results by construction, we employed such a strategy to determine the simultaneous (common) description of the same experimental  data by two different approaches. In all scenarios of CFO studied here
we, indeed, always  found the region where these two approaches provide a simultaneous and good description of the data. 
Such a strategy allows us to get rid of the  existing ambiguity in the light (anti)nuclei data description and to determine the 
CFO parameters of nuclei in A+A collisions of high energy with high confidence. 
In particular, for the ALICE data measured at 
$\sqrt{s_{NN}} =2.76$ TeV  we found that the separate CFO of nuclei provides  a very high   accuracy in the description of hadronic multiplicity ratios and the light (anti)nuclei yields using only 3 fitting parameters with $\chi^2_{tot}/dof \simeq 0.758$.
The found CFO  temperature of nuclei is   $T_A^{com}= 175.1^{+2.3}_{-3.9}$ MeV and their CFO volume is $V_A^{com} = 2660^{+1010}_{-1160}$ fm$^3$, while the CFO of hadrons occurs at essentially lower temperature.

On the contrary, from the analysis of the STAR data measured at $\sqrt{s_{NN}} =200$ GeV, we found that the single CFO of hadrons and nuclei with 3 fitting parameters provides a better description which, in addition,  is internally self-consistent.  
In this case the best description of the STAR data  is achieved for the following CFO parameters of nuclei:   $T_A^{com} = T_h = 167.2 \pm 3.9$ MeV,  $V_A^{com} = 1898.5  \pm 157.5$ fm$^3$  and $\chi^2_{tot}/dof \simeq 1.45$.

Based on the idea that the light (anti)nuclei are produced from the quark-gluon bags with an exponential mass spectrum, 
we interpret the found CFO volumes of nuclei as the sum of volumes of  quark-gluon bags. From this interpretation,  
we estimated the ratio of initial entropy of thermalized bags for A+A collision energies $\sqrt{s_{NN}}$ $=200$ GeV and $\sqrt{s_{NN}} =2.76$ TeV 
and the number of the emitting sources of nuclei, which, in principle,  can be verified by the hydrodynamic or hydro-kinetic approaches.   
Surprisingly, if the number of such sources is 14 for the ALICE energy and, consequently,  10 for the STAR energy,
as it is required by their common CFO volumes, then the radius of emitting sources of nuclei is 3.566 fm, which practically  coincides with the  value of the coalescence distance used in a successful transport code simulating the production of nuclei \cite{Coalesc1}. 

{ In the present work we demonstrate that the experimental data for the yields of hadrons and light nuclei produced in heavy-ion collisions at 
RHIC and LHC energies can be described with a very high accuracy, if one uses a formulation  of the HRGM that employs the classical 
second virial coefficients corresponding to a hard-sphere model of nuclei and hadrons.
At the same time it is shown that the determination of  the CFO temperature of light nuclei is a rather delicate issue since the result depends on the underlying scenario of their CFO.  

The simple model of hard spheres for the repulsive interactions between hadrons and nuclei as employed in the present work can of course only be considered 
as an intermediate step in our understanding of the formation of hadrons and light nuclei from a hadronizing quark-gluon plasma. 
Microphysical approaches should be further developed which treat hadrons and nuclei as multi-quark clusters and would allow for a deeper understanding of the hadrochemistry on the quark level. 
One aspect of a description on this level would be the explanation of the short-range repulsion by quark Pauli blocking among hadrons (see, e.g., Ref.~\cite{Blaschke:2020qrs}), eventually augmented by repulsive multi-pomeron exchange forces that have proven essential to describe large-angle nucleus-nucleus scattering at the Fermi energy and in resolving the hyperon puzzle of neutron star structure  \cite{Yamamoto:2015lwa}.   
From the further systematic analysis of light nuclei production measured in the heavy ion collision experiments a picture may emerge in which 
the puzzling result of a high CFO temperature finds its explanation by a hadronization of multi-quark states from the QGP, 
as it was already anticipated in Ref. \cite{Chapline:1978kg} and emphasized again in Ref.  \cite{PBM18} in the light of the recent experiments.
}
\\

\noindent
{\bf Author contributions.}
{
K.A.B. developed the idea behind this work and together with D.B.B. took the lead in writing the manuscript.
O.V.V., B.E.G., V.V.S. and E.S.Z. performed fit of the experimental data on the light (anti)nuclei and hadrons.
O.I.I., N.S.Y.  and E.G.N. verified the analytical methods. Both N.S.Y. and  S.V.K., helped in calculating the CFO volume of hadrons and designed the figures. G.M.Z., L.V.B., E.E.Z., S.K., G.R.F. and A.V.T. contributed to the interpretation of the results and provided a critical feedback. All authors discussed the results and contributed to the final manuscript.\\
  }

\noindent
{\bf Acknowledgments.}
The authors are thankful to Dmytro Oliinychenko  for brining  to our attention   Ref.  \cite{STARA3} and for illuminating discussions, and  to Grigory Nigmatkulov and   Ivan Yakimenko  for  the valuable comments. 
K.A.B.  and G.M.Z. acknowledge support from the NAS  of Ukraine by its priority project ``Fundamental properties of the matter in the relativistic collisions of nuclei and in the early Universe"
(No. 0120U100935).
V.V.S. and  O.I.I. are thankful for the support by the Funda\c c\~ao para a Ci\^encia e Tecnologia (FCT), Portugal, by the project UID/04564/2020. 
The work of O.I.I. was supported by  the project CENTRO-01-0145-FEDER-000014 via 
the CENTRO 2020 program, and  POCI-01-0145-FEDER-029912 with financial support from POCI, in its FEDER component and by the FCT/ MCTES budget via national funds (OE).
The work of L.V.B. and E.E.Z. was supported by the Norwegian Research Council (NFR) under grant No. 255253/ F53 CERN Heavy Ion Theory, and by the RFBR grants 18-02-40085 and 18-02-40084. 
K.A.B., O.V.V.,  N.S.Ya. and  L.V.B.  thank the Norwegian Agency for International Cooperation and Quality Enhancement in Higher Education for the financial support under grants  CPEA-LT-2016/10094 and UTF-2016-long-term/10076.
A.V.T. acknowledges partial support from RFBR under grant No.
18-02-40086 and from the Ministry of Science and Higher Education of the Russian Federation,
Project ``Fundamental properties of elementary particles and cosmology" No 0723-2020-0041.
D.B.B. received funding from the RFBR under  grant No. 18-02-40137.
D.B.B. and A.V.T.  acknowledge
 a  partial support from the National Research Nuclear University ``MEPhI'' in the framework of the Russian Academic Excellence Project (contract no. 02.a03.21.0005, 27. 08.2013).  
The authors  are  grateful to the COST Action CA15213 ``THOR" for supporting their  networking.


\begin{thebibliography}{99}

\bibitem{Ebeling:2008mg} 
  W.~Ebeling, D.~Blaschke, R.~Redmer, H.~Reinholz and G.~R\"opke,
  J.\ Phys.\ A {\bf 42}, 214033 (2009).

\bibitem{Ropke:2018ewt} 
  G.~R{\"o}pke, D.~Blaschke, T.~D{\"o}ppner, C.~Lin, W.~D.~Kraeft, R.~Redmer and H.~Reinholz,
  Phys.\ Rev.\ E {\bf 99}, no. 3, 033201 (2019).

\bibitem{Blaschke:2020qrs}
D.~Blaschke, H.~Grigorian and G.~R\"opke,
Particles \textbf{3}, no.2, 477-499 (2020).

\bibitem{Typel:2009sy} 
  S.~Typel, G.~R\"opke, T.~Kl\"ahn, D.~Blaschke and H.~H.~Wolter,
  Phys.\ Rev.\ C {\bf 81}, 015803 (2010).

\bibitem{Hempel:2015yma} 
  M.~Hempel, K.~Hagel, J.~Natowitz, G.~R{\"o}pke and S.~Typel,
  Phys.\ Rev.\ C {\bf 91}, no. 4, 045805 (2015).

\bibitem{Ropke:2020peo}
G.~R\"opke,
Phys. Rev. C \textbf{101}, no.6, 064310 (2020).
  
\bibitem{Lattimer:1991nc} 
  J.~M.~Lattimer and F.~D.~Swesty,
  Nucl.\ Phys.\ A {\bf 535}, 331 (1991).
  
\bibitem{Shen:1998gq} 
  H.~Shen, H.~Toki, K.~Oyamatsu and K.~Sumiyoshi,
  Nucl.\ Phys.\ A {\bf 637}, 435 (1998).


\bibitem{Hempel:2011kh} 
  M.~Hempel, J.~Schaffner-Bielich, S.~Typel and G.~R\"opke,
  Phys.\ Rev.\ C {\bf 84}, 055804 (2011).
  
\bibitem{Ropke:2012qv} 
  G.~R\"opke, N.-U.~Bastian, D.~Blaschke, T.~Kl\"ahn, S.~Typel and H.~H.~Wolter,
  Nucl.\ Phys.\ A {\bf 897}, 70 (2013); 
  [arXiv:1209.0212 [nucl-th]].

\bibitem{KAB_Phi-approach3} 
N.~U.~F.~Bastian, D.~Blaschke, T.~Fischer and G.~R{\"o}pke,
  Universe {\bf  4}, 67 (2018) and references therein.

 {
\bibitem{Mrowczynski:2016xqm}
S.~Mrowczynski,
Acta Phys. Polon. B \textbf{48}, 707 (2017).

\bibitem{Sun:2017xrx}
K.~J.~Sun, L.~W.~Chen, C.~M.~Ko and Z.~Xu,
Phys. Lett. B \textbf{774}, 103-107 (2017).

\bibitem{Sun:2018jhg}
K.~J.~Sun, L.~W.~Chen, C.~M.~Ko, J.~Pu and Z.~Xu,
Phys. Lett. B \textbf{781}, 499-504 (2018).

\bibitem{Sun:2018mqq}
K.~J.~Sun, C.~M.~Ko and B.~D\"onigus,
Phys. Lett. B \textbf{792}, 132-137 (2019).

\bibitem{Vovchenko:2018fiy}
V.~Vovchenko, B.~D\"onigus and H.~Stoecker,
Phys. Lett. B \textbf{785}, 171-174 (2018).

\bibitem{Bellini:2018epz}
F.~Bellini and A.~P.~Kalweit,
Phys. Rev. C \textbf{99}, no.5, 054905 (2019).

\bibitem{Bellini:2020cbj}
F.~Bellini, K.~Blum, A.~P.~Kalweit and M.~Puccio,
[arXiv:2007.01750 [nucl-th]].

\bibitem{Cai:2019jtk}
Y.~Cai, T.~D.~Cohen, B.~A.~Gelman and Y.~Yamauchi,
Phys. Rev. C \textbf{100}, no.2, 024911 (2019).

\bibitem{Aichelin:2019tnk}
J.~Aichelin, E.~Bratkovskaya, A.~Le F\`evre, V.~Kireyeu, V.~Kolesnikov, Y.~Leifels, V.~Voronyuk and G.~Coci,
Phys. Rev. C \textbf{101}, no.4, 044905 (2020).

	\bibitem{Vitiyuk_Ref2}
	%
	D. Oliinychenko, 
	talk given  at XXVIIIth  Conference ``Quark Matter 2019``, 
	arXiv:2003.05476v1 [hep-ph] and references therein.

	
	
	\bibitem{Vitiyuk_Ref3}
	%
	S. Mrowczynski,
	arXiv:2004.07029v1 [nucl-th] and references therein.



}

\bibitem{Blaschke:2020jmv} 
  D.~Blaschke, A.~V.~Friesen, Y.~B.~Ivanov, Y.~L.~Kalinovsky, M.~Kozhevnikova, S.~Liebing, A.~Radzhabov and 
  G.~R{\"o}pke,
  arXiv:2004.01159 [hep-ph].
  
\bibitem{Blaschke:2020gqr} 
  D.~Blaschke, G.~R{\"o}pke, Y.~Ivanov, M.~Kozhevnikova and S.~Liebing,
  Springer Proc. Phys. {\bf 250}, 183 (2020).


\bibitem{MHRGM1}
%
D. R. Oliinychenko, K. A. Bugaev and A. S. Sorin, 
Ukr. J. Phys.  {\bf 58},  211  (2013).

  
\bibitem{MHRGM2}
%
K. A. Bugaev, D. R. Oliinychenko,   A. S. Sorin and G. M. Zinovjev, 
Eur. Phys. J. A {\bf 49}, 30  (2013).


\bibitem{MHRGM3}
%
 K. A.~Bugaev {\it et al}., 
 Europhys. Lett. {\bf 104},   (2013) 22002.
  
 \bibitem{MHRGM4}
%
  K. A. Bugaev,  A. I. Ivanytskyi,  D. R. Oliinychenko, E. G. Nikonov, V. V. Sagun   and G. M. Zinovjev,
Ukr. J. Phys. {\bf 60},  181 (2015). 
  
 \bibitem{Sagun14}
%
V. V. Sagun, 
{Ukr. J. Phys.}  {\bf 59}, 755  (2014).


\bibitem{GSA15}
%
 K. A.~Bugaev {\it et al}., 
{Phys. Part. Nucl. Lett.} {\bf 12}, 238 (2015).

\bibitem{GSA16}
%
K. A. Bugaev {\it et al}., 
{Eur. Phys. J. A} {\bf  52}, 175 (2016). 

\bibitem{GSA16b}
%
K. A. Bugaev {\it et al}., 
{Eur. Phys. J. A} {\bf  52}, 227 (2016). 

\bibitem{Signals18}
%
 K. A. Bugaev {\it et al.},
{Phys. Part. Nucl. Lett.} {\bf 15},  210 (2018).


\bibitem{Signals19}
%
K. A. Bugaev {\it et al}., 
{EPJ Web of Conferences} {\bf 204}, 03001  (2019).


\bibitem{PBM06}
%
 A. Andronic, P.Braun-Munzinger and  J. Stachel,
 {Nucl. Phys. A } {\bf 772},  167 (2006) and references therein.


	\bibitem{IST1}
	%
V. V. Sagun, A. I. Ivanytskyi, K. A. Bugaev and I. N. Mishustin,
Nucl. Phys. A {\bf 924},  24  (2014).

	\bibitem{IST2}
	%
	V. V. Sagun et al., 
 Eur. Phys. J. A {\bf 54},   100 (2018).

	\bibitem{IST3}
	%
K. A. Bugaev et al., 
Nucl. Phys. A {\bf 970},  133 (2018).

 
\bibitem{QSTAT2019}	
%
K. A. Bugaev, 
 Eur. Phys. J. A {\bf 55},  215 (2019).

\bibitem{Nazar2019}
%
N. S. Yakovenko, K. A. Bugaev, L.V. Bravina and E. E. Zabrodin,
  arXiv:1910.04889 [nucl-th] p. 1-13.
  
\bibitem{Bazak:2020wjn}
S.~Bazak and S.~Mrowczynski,
Eur. Phys. J. A \textbf{56}, no.7, 193 (2020).
 

\bibitem{STARA1}
%
STAR Collaboration (B. I. Abelev {\it et al}.),
Science {\bf 328}, No 5974, p. 58-62 (2010).

\bibitem{STARA2}
%
STAR Collaboration (H. Agakishiev {\it et al}.),
 Nature  {\bf 473},  No 7347,  p. 353-356 (2011).
 
 \bibitem{STARA3}
%
STAR Collaboration (J. Adam {\it et al}.), 
Phys. Rev. C  {\bf 99}, 064905 (2019).
  
\bibitem{KAB_Ref1a}
%
ALICE Collaboration (J. Adam {\it et al}.),  { Phys. Rev. C } {\bf 93},  024917  (2016).

\bibitem{KAB_Ref1b}
%
ALICE Collaboration (L. Ramona{\it et al}.), 
{AIP Conf. Proc.} {\bf 1701},  (1)  080009 (2016).


\bibitem{KAB_Ref1c}
%
ALICE Collaboration (J. Adam {\it et al}.), 
{Phys.\ Lett.\ B} {\bf 754},  360 (2016).
  
 {
\bibitem{Raju}
%
{R. Venugopalan and M. Prakash}, 
{Nucl. Phys. A} {\bf  546}, {718} (1992).    

\bibitem{Edward2018}
%
E.~Shuryak and J.~M.~Torres-Rincon,
Phys. Rev. C \textbf{100}, 024903   (2019) and references therein. 

\bibitem{Shuryak:2019ikv}
E.~Shuryak and J.~M.~Torres-Rincon,
Phys. Rev. C \textbf{101}, no.3, 034914 (2020).

\bibitem{Satarov10}
%
L. M. Satarov,  M. N. Dmitriev and I. N. Mishustin,
Phys. Atom. Nucl.  {\bf  72},  1390  (2009).

\bibitem{ISTQ}
%
 K. A. Bugaev, A. I. Ivanytskyi, V. V. Sagun, E. G. Nikonov and G. M. Zinovjev,
Ukr. J. Phys. {\bf  63},  863  (2018) 
and references therein

\bibitem{KABkinFO1}
%
K. A. Bugaev,
Nucl. Phys. A {\bf  606}, 559 (1996).

\bibitem{KABkinFO2}
%
K. A. Bugaev,
Phys. Rev. Lett. {\bf 90}, 252301 (2003) and references therein

\bibitem{KAB_Ref27}
%
R. Hagedorn, 
  {Nuovo Cim. Suppl.}
{\bf 3},  147 (1965).
}
 
 \bibitem{KAB_Chatter15}
%
S. Chatterjee {\it et al.,}
{Adv. High Energy Phys.}  2015, 349013
(2015) and references therein.
 
 \bibitem{KAB_Jean}
%
J. Cleymans, S. Kabana, I. Kraus, H. Oeschler,  K. Redlich and N. Sharma,
{Phys. Rev. C}  {\bf 84} 054916 (2011).   

\bibitem{KAB_Ref2}
%
J. Stachel, A. Andronic, P. Braun-Munzinger  and K.  Redlich, 
 {J.\ Phys.\ Conf.\ Ser. }  {\bf 509},  012019 (2014).

\bibitem{PBM18}
A.~Andronic, P.~Braun-Munzinger, K.~Redlich and J.~Stachel,
Nature \textbf{561}, no.7723, 321-330 (2018).
%

\bibitem{KAB_Ref3}
%
K. A. Bugaev {\it et al}., 
{J. of Phys. Conf. Series}  {\bf 1390},  012038  (2019).
 

  
\bibitem{PBM19}
 %
P. Braun-Munzinger  and  B. D{\"o}nigus,
{Nucl. Phys. A} {\bf 987}, 144  (2019) and references therein.

{
	
}

\bibitem{Grinyuk2020}
%
B. E. Grinyuk {\it et al}.,
arXiv:2004.05481v1 [hep-ph] (2020).





\bibitem{KAB_Bohr}
%
A. Bohr  and B. Mottelson,   {\it Nuclear Structure}, vol. 1 (Benjamin, New York, 1969).

\bibitem{RecentRms}
%
I. Angeli and K. Marinova, 
At. Data Nucl. Data Tables {\bf 99}, 69 (2013). 

\bibitem{HtritonR}
%
H. Nemura, Y. Suzuki, Y. Fujiwara, C. Nakamoto, 
Prog. Theor. Phys. {\bf 103}, 929   (2000); 
arXiv:nucl-th/9912065.

\bibitem{Rafelski}
J. Rafelski, Phys. Lett. B {\bf 62}, 333 (1991).

\bibitem{Beth:1937zz} 
E.~Beth and G.~Uhlenbeck,
  Physica {\bf   4}, 915 (1937).

\bibitem{Hufner:1994ma} 
  J.~H\"ufner, S.~P.~Klevansky, P.~Zhuang and H.~Voss,
  Annals Phys.\  {\bf 234}, 225 (1994).

\bibitem{Wergieluk:2012gd} 
  A.~Wergieluk, D.~Blaschke, Y.~L.~Kalinovsky and A.~Friesen,
  Phys.\ Part.\ Nucl.\ Lett.\  {\bf 10}, 660 (2013).

\bibitem{Blaschke:2013zaa} 
  D.~Blaschke, M.~Buballa, A.~Dubinin, G.~R\"opke and D.~Zablocki,
  Annals Phys.\  {\bf 348}, 228 (2014).


\bibitem{Dubinin:2016wvt} 
  D.~Blaschke, A.~Dubinin, A.~Radzhabov and A.~Wergieluk,
  Phys.\ Rev.\ D {\bf 96}, no. 9, 094008 (2017).



\bibitem{KAB_David:16A}
 %
D. Blaschke, A. Dubinin and L. Turko, arXiv:1611.09845v2 [hep-ph].
 

 \bibitem{KAB_David:16B}
 %
D.~Blaschke, A.~Dubinin and L.~Turko,
 Acta Phys.\ Polon.\ Supp.\  {\bf  10}, 473 (2017).

 \bibitem{KAB_Phi-approach}
%
G. ~Baym, Phys. Rev. {\bf  127}, 1391 (1962).

 \bibitem{KAB_Phi-approach2}
%
 B. ~Vanderheyden and  G. ~Baym, J. Stat. Phys.  {\bf  93}, 843 (1998).

\bibitem{Reuter08}
	%
	K. A. Bugaev and P. T. Reuter, 
	 Ukr. J. Phys. {\bf 52},  489  (2007) and references therein. 

\bibitem{Huang}
	K. Huang, {\it Statistical Mechanics} (Wiley \& Sons, New York, 1967)
	
\bibitem{Satarov_Cs}
%
L. M. Satarov, K. A. Bugaev, I. N. Mishustin, 
Phys. Rev. C {\bf  91},   055203 (2015).

\bibitem{Vovchenko-ALICE}
 V. Vovchenko, H. St{\"o}cker, 
 J. Phys. G  {\bf 44},   055103 (2017).
 
 \bibitem{Simple-Liquids}
%
J. P. Hansen and I. R. McDonald, {\it  Theory of Simple Fluids} (Academic Press, Amsterdam, 2006).

\bibitem{MITBagM}
%
A. Chodos, R. L. Jaffe, K. Johnson, C. B. Thorn, V. F. Weisskopf, 
Phys. Rev. {\bf D 9},    3471  (1974). 


{
 
 \bibitem{Abelev:2013vea}
  B.~Abelev {\it et al.} [ALICE Collaboration],
  {Phys.\ Rev.\ C} {\bf 88} (2013) 044910.

\bibitem{Abelev:2013zaa}
  B.~B.~Abelev {\it et al.} [ALICE Collaboration],
 {Phys.\ Lett.\ B} {\bf 728} (2014) 216;
  Erratum: [{Phys.\ Lett.\ B} {\bf 734} (2014) 409].

\bibitem{Abelev:2013xaa}
  B.~B.~Abelev {\it et al.} [ALICE Collaboration],
 {Phys.\ Rev.\ Lett.}\  {\bf 111} (2013) 222301.

\bibitem{Abelev:2015prc}
   B.~B.~Abelev {\it et al.} [ALICE Collaboration],
  {Phys.\ Rev.\ C} {\bf 91} (2015)  024609.
}


\bibitem{KAB_pbm17}
%
A. Andronic, P. Braun-Munzinger, K. Redlich and J. Stachel,
{J. Phys. Conf. Ser.} {\bf 779},  012012   (2017).

\bibitem{KAB_lqcd1}
%
Wuppertal-Budapest Collaboration
(S. Borsanyi  {\it et al}.),  {\it JHEP}
{\bf 1009},  073 (2010).

\bibitem{KAB_lqcd2}
%
HotQCD Collaboration
(A. Bazavov {\it et al}.), 
{Phys. Rev. D} {\bf  90},  094503 (2014). 

{
\bibitem{Bazavov:2018mes}
A.~Bazavov \textit{et al.} [HotQCD],
Phys. Lett. B \textbf{795} (2019), 15-21.
}

{ 
\bibitem{KABstar:200a}
%
J.~Adams {\it et al.,}
Phys. Rev. Lett. {\bf 92}, 112301 (2004).

\bibitem{KABstar:200b}
%
J.~Adams {\it et al.,}
Phys. Lett. B {\bf 612},  181 (2005).

\bibitem{KABstar:200c}
%
A.~Billmeier {\it et al.,}
J. Phys. G {\bf 30}, S363 (2004).
}

\bibitem{PBM2010}
%
A. Andronic, P. Braun-Munzinger, J. Stachel and H. Stoecker,
 Phys. Lett. B {\bf 697},  203 (2011). 

\bibitem{Rapp2018}
%
X. Xu and R. Rapp, 
Eur. Phys. J. A \textbf{55},  68  (2019); 
arXiv:1809.04024v2 [nucl-th] and references therein.

{
\bibitem{Chapline:1978kg}
G.~F.~Chapline and A.~K.~Kerman,
MIT-CTP-695 (1978).}

\bibitem{Thermostat1}
%
L. G. Moretto,  K. A. Bugaev, J. B. Elliott and L. Phair,
{Europhys. Lett. } {\bf 76},  402 (2006); 
LBNL preprint {56898}.

{
\bibitem{Gallmeister:2020fiv}
K.~Gallmeister and C.~Greiner,
[arXiv:2007.08258 [hep-ph]].
}

\bibitem{IHKm}
%
V. Yu. Naboka, Iu. A. Karpenko and Yu. M. Sinyukov, Phys. Rev. C {\bf 93},   024902 (2016). 


\bibitem{Coalesc1}
%
S. Sombun {\it  et al.}, Phys. Rev. C {\bf 99}, 014901 (2019).

{
\bibitem{Yamamoto:2015lwa}
Y.~Yamamoto, T.~Furumoto, N.~Yasutake and T.~A.~Rijken,
Eur. Phys. J. A \textbf{52}, no.2, 19 (2016).
}



\end{thebibliography}
\end{document}